\documentclass[times,doublespace,useAMS,usenatbib]{qjrms4}
\usepackage{bm}

\usepackage{booktabs} % To thicken table lines
\usepackage[colorlinks,bookmarksopen,bookmarksnumbered,citecolor=red,urlcolor=red]{hyperref}
\usepackage{moreverb}
\usepackage{cancel}
\usepackage{array}
\newcolumntype{P}[1]{>{\centering\arraybackslash}p{#1}}
\newcolumntype{M}[1]{>{\centering\arraybackslash}m{#1}}

\usepackage[normalem]{ulem} % strike text through
\definecolor{rred}{rgb}{0.7,0.,0.}
\definecolor{ggreen}{rgb}{0.,0.5,0.}

\newcommand{\x}{\bf x}
\newcommand{\y}{\bf y}

\usepackage{xcolor}

\def\be{\begin{equation}}
\def\ee{\end{equation}}
\def\bea{\begin{eqnarray}}
\def\eea{\end{eqnarray}}
\def\nn{\nonumber\\}
\def\({\left(}
\def\){\right)}

\def\x{\mathbf{x}}
\def\barx{\overline{\mathbf x}}

\def\y{\mathbf{y}}
\def\w{\mathbf{w}}

\def\bA{\mathbf{A}}

\def\bR{\mathbf{R}}
\def\bP{\mathbf{P}}

\def\bH{\mathbf{H}}
\def\bI{\mathbf{I}}

\def\bX{\mathbf{X}}
\def\bY{\mathbf{Y}}

\def\bZ{\mathbf{Z}}
\def\bU{\mathbf{U}}
\def\bV{\mathbf{V}}
\def\bS{\mathbf{S}}

\def\eeta{{\bm \eta}}
\def\epsi{{\bm \epsilon}}
\def\bchi{{\bm \chi}}

\def\bzero{{\mathbf 0}}
\def\b1{{\mathbf 1}}
\def\T{\mathrm{T}}

\def\volumeyear{2016}

\begin{document}

\runningheads{A. Carrassi, M. Bocquet, A. Hannart and M. Ghil}{ Data assimilation for model evidence }

\title{Estimating model evidence using data assimilation}

\author{A.~Carrassi\affil{a}\corrauth,  M. Bocquet\affil{b}, A. Hannart\affil{c} and M. Ghil\affil{d,e}}

\address{\affilnum{a} Nansen Environmental and Remote Sensing Center, Bergen, Norway. \\
\affilnum{b} CEREA, joint laboratory \'Ecole des Ponts ParisTech and EDF R\&D, Universit\'e Paris-Est, Champs-sur-Marne, France. \\
\affilnum{c} IFAECI, CNRS-CONICET-UBA, Buenos Aires, Argentina. \\
\affilnum{d} \'Ecole Normale Sup\'erieure, Paris, France.\\
\affilnum{e} University of California, Los Angeles, USA}

\corraddr{Alberto Carrassi,  e-mail: alberto.carrassi@nersc.no }

%\history{Manuscript received xx xxxx xx; in final form xx
%xxxx xx}

\begin{abstract}
\begin{spacing}{1.2}
We review the field of data assimilation (DA) from a Bayesian perspective and show that, in addition to its by now common application to state estimation, DA may be used for model selection. An important special case of the latter is the discrimination between a factual model --- which corresponds, to the best of the modeler's knowledge, to the situation in the actual world in which a sequence of events has occurred --- and a counterfactual model, in which a particular forcing or process might be absent or just quantitatively different from the actual world. Three different ensemble-DA methods are reviewed for this purpose: the ensemble Kalman filter (EnKF), the ensemble four-dimensional variational smoother (En-4D-Var), and the iterative ensemble Kalman smoother (IEnKS). An original contextual formulation of model evidence (CME) is introduced. It is shown how to apply these three methods to compute CME, using the approximated time-dependent probability distribution functions (pdfs) each of them provide in the process of state estimation. The theoretical formulae so derived are applied to two simplified nonlinear and chaotic models: (i) the Lorenz three-variable convection model (L63), and (ii) the Lorenz 40-variable mid-latitude atmospheric dynamics model (L95). The numerical results of these three DA-based methods and those of an integration based on importance sampling are compared. It is found that better CME estimates are obtained by using DA, and the IEnKS method appears to be best among the DA methods. Differences among the performance of the three DA-based methods are discussed as a function of model properties. Finally, the methodology is implemented for parameter estimation and for event attribution.
\end{spacing}
\end{abstract}


\maketitle

\section{Introduction and motivation} \label{sec:intro}

High-dimensional and nonlinear state-space evolution models arise in numerous and diverse fields, {\it e.g.,} numerical weather  prediction (NWP) \citep{GMR91}, air quality forecasting \citep{zhang2012}, subsurface flow modeling
\citep{Els14a,Els14b}, oceanography \citep{Bal09}, climatic projection and reconstruction \citep{Bhe12}, and signal
processing \citep{Cris02}, to mention just a few. Most of these applications require one to solve the problem of optimally estimating the partially observed state that evolves in time. This problem is often tackled by combining the observed data 
with a numerical model that is based on a comprehensive theoretical description of the processes at play. Designing methods and numerical algorithms that address this problem in a way that is theoretically sound and yet tailored to the 
specifics of the context at hand is an active research field in computational statistics in general, and in the aforementioned fields in particular.

This paper focuses on a related yet distinct problem, which has received less attention thus far. It consists in
quantifying the resulting performance of the 
state inference by estimating the so-called {\em marginal
likelihood of the observations} --- also  
referred to as {\em model evidence} -- which quantifies the ``goodness-of-fit''
between the data and the chosen state-space model \citep{Baum70}. Solving the model evidence estimation problem has attracted, as we shall see, less research effort than the state estimation one. 
Still, estimating model evidence is just as
important as state estimation. 

Indeed, the model evidence can be used as a general metric for model comparison and selection. Areas of application include calibrating a given state-space model's parameters based on the 
observed data by maximizing the model evidence;
comparing the skill of several candidate models (or model settings, or boundary conditions, or numerical schemes) in representing a given
observed phenomenon or class of phenomena and thereby selecting the most appropriate candidate; quantifying the evidence that supports one of several, potentially conflicting, theoretical hypotheses for the physics of a phenomenon; or providing evidence for either the existence or the nonexistence of a causal relationship between a hypothetical forcing external to a system and an observed response of the system. 

It is, in fact, the latter situation that largely motivated the present work, in the context of the climate system 
and of its increasingly significant anthropogenic forcing, cf.~\citet{Hannart-et-al-2016} and references therein. More
generally speaking, model evidence may help address fundamental questions whenever
confronting theory to observations in an attempt to improve the former, and prove to be a useful tool in extracting quantitative information from such a model--data confrontation.

Mathematically speaking, deriving model evidence is quite difficult as it requires, by definition,
to integrate out the state vector. Although viable solutions exist in the case of Gaussian errors and linear model \citep[see e.g.][]{Winiarek-et-al-2011}, this daunting task is usually intractable, especially for the class of high-dimensional and non-linear 
models considered here. To circumvent this difficulty, different approaches were proposed in the literature to
yield an accurate and more easily computable estimate of the desired value \citep{HK01,Pitt02,Kan09}. 

Recently, marginal likelihoods have been used for the
purposes of Bayesian inference and model selection. An efficient approach to deal with the former, based on Monte Carlo
sampling, has been proposed by \citet{Els14a,Els14b} in the context of models of subsurface flows. 
\citet{Carson_et_al_2015} have described an application to model selection; the method is again based on an advanced
Monte Carlo technique and has proven to efficiently discriminate between phenomenological models of the
glacial-interglacial cycle using a single dataset.
\citet[][Chapter 9.1]{Reich-Cotter-2015} offer an exposition on model evidence, its use for model selection and how to use particle filters for its computation.

The present study focuses on the estimation of model evidence using
data assimilation (DA) methods designed to deal with large numerical models and datasets, subject to partially Gaussian assumptions.  
DA methods were in fact initially developed for high-dimensional state
estimation in the NWP context,  
in order to initialize an atmospheric model by estimating
its state variables based on meteorological observations that are incomplete, diverse, unevenly distributed in space and
time and are contaminated by measurement error \citep[and references therein]{Bengtsson81, Kalnay2002}. 
These methods outgrew their original application field over the past decades, to reach a wide variety of fields in the geosciences and elsewhere. They thus present the advantage of being used operationally in many different contexts and by many practitioners, and to be the focus of a
significant research effort \citep{blayo2015}. In the recent past, model evidence estimates have started to appear in the DA context \citep{Winiarek-et-al-2011,winiarek2012,Tand15,Hannart-et-al-2016}.

The purpose of the present paper is to extend these studies by proposing several approaches for estimating the marginal likelihood (as model evidence). 
Section \ref{sec:DAB} recalls key aspects of Bayesian DA that are essential for the discussion; see \cite{Bocquet2010} and \cite{Reich-Cotter-2015} for an extensive treatment of the subject. 
Section \ref{sec:evidence} introduces the concept of {\em contextual} model evidence which is of particular interest in the DA setting. Section \ref{sec:DAM} lays out four
different computational schemes to estimate contextual model evidence. Section \ref{sec:results} implements these schemes for two low-dimensional nonlinear models and compares their performance. 
Section \ref{sec:apply} briefly illustrates the application of the proposed schemes to parameter estimation and causal attribution of weather events. Section \ref{sec:concl} finally summarizes our results, provides conclusions and future directions.

\section{Data assimilation from a Bayesian perspective: brief overview \label{sec:DAB}}

Let us assume that a model of the physical process of interest is given as a discrete dynamical system in an $M$-dimensional Euclidean space $\mathbb R^M$,
\begin{equation}
\label{model}
\x_{k} = \mathcal{M}_{k:k-1}(\x_{k-1}) + \eeta_{k}.
\end{equation}
Here $\x_{k}\in{\mathbb R}^M$ is the state vector, $\mathcal{M}_{k:k-1}: {\mathbb R}^M \rightarrow {\mathbb R}^M$ is usually 
a nonlinear, possibly chaotic, map and $\eeta_k\in{\mathbb R}^M$ stands for model error, represented as a stochastic additive term.  

Noisy observations of $\x$ are available at discrete times and are represented as components of the observation vector ${\bf y}\in{\mathbb R}^d$. 
The relation between $\y$ and the model state $\x$ is given by
\begin{equation}
\label{obs}
{\bf y}_k={\mathcal H}_k(\x_k)+\epsi_k.
\end{equation}
In Eq.~\eqref{obs}, ${\mathcal H}:{\mathbb R}^M \rightarrow {\mathbb R}^d$ is the, possibly nonlinear, observation operator that maps the model solution to the observation space  ${\mathbb R}^d$; typically $d \ll M$ and ${\mathcal H}$ may involve spatial interpolations in finite-difference models or spectral-to-physical space transformation in spectral models. 
Transformations based on physical laws for indirect measurements, such as radiative fluxes used to infer temperatures, can also be represented in this way \citep[{e.g.,}][]{Kalnay2002}.
The observational error ${\bm\epsilon}_k$ is also represented as a stochastic additive term. 

Both random sequences $\{\eeta_k : k=0, \ldots ,K\}$ and $\{\epsi_k : k=0, \ldots, K\}$ are assumed to be white in time, mutually independent and distributed according to the probability density functions (pdfs) $p_{\eeta}$ and $p_{\epsi}$, respectively. These pdfs represent the transitional kernel for the probabilistic transition from $\x_{k-1}$ to $\x_{k}$, and the likelihood of the observations $\y_k$ conditioned on the state $\x_k$, respectively,
\begin{align}\label{eq:trans}
p(\x_{k}\vert\x_{k-1})& = p_{\eeta}[\x_{k} - \mathcal{M}_k(\x_{k-1})] , \\
p(\y_{k}\vert\x_{k})& = p_{\epsi}[\y_{k} - \mathcal{H}_k(\x_k)].
\end{align} 

The output of the estimation process is the posterior pdf $p(\x\vert\y)$ of the process $\x$ conditioned on the data $\y$, 
\begin{equation}
\label{Bayes-Th}
p(\x\vert\y)=\frac{p(\y\vert\x)p(\x)}{p(\y)}\, .
\end{equation}
In this straightforward application of the Bayes formula, $p(\x)$ is the prior distribution that encodes all the knowledge about the process before assimilating the new observations, and $p(\y)$ is the observation likelihood. The latter is usually independent from time when the estimation is performed and plays the role of a normalization coefficient. We will see in sections \ref{sec:evidence} and \ref{sec:DAM} how this view can be reversed when solving the DA problem for the purpose of evaluating model evidence, cf. \cite{Hannart-et-al-2016}.  

Once the sequences of system states and observations are collected into $\x_{k:0}=\{\x_k,\x_{k-1},...,\x_0\}$ and $\y_{k:0}=\{\y_k,\y_{k-1},...,\y_0\}$, it is possible to define three estimation problems, depending on the time period where observations are distributed and the time when we want to estimate the state \citep{Wiener49}:
\begin{enumerate}
\item {\it Prediction}: Estimate $p(\x_l\vert\y_{k:0})$ with $l>k$.
\item {\it Filtering}: Estimate $p(\x_k\vert\y_{k:0})$.
\item {\it Smoothing}: Estimate $p(\x_{k:0}\vert\y_{k:0})$. 
\end{enumerate}
The prediction problem is formally addressed by solving the corresponding Chapman-Kolmogorov equation for the propagation of a pdf under the model dynamics
\begin{equation} 
\label{Chap-Kol}
p(\x_l\vert\y_{k:0}) = \int \! \mathrm{d}\x_k \, p_{\eeta}[\x_l - \mathcal{M}_{l:k}(\x_{k})]p(\x_k\vert\y_{k:0}).
\end{equation} 
The filtering problem is the most common one in geophysical applications, and it is characterized by sequential processing, in which measurements are utilized as they become available \citep{jazwinski1970,Bengtsson81}:
a so-called analysis step, in which the conditional pdf $p(\x_k\vert\y_{k:0})$ is updated using the latest observation, $\y_k$, alternates with a forecast step in which this pdf is propagated forward until the time of a new observation. The analysis is based on the application of the Bayes formula \eqref{Bayes-Th}, which 
becomes
\begin{equation}
\label{filt-Bay}
p(\x_k\vert\y_{k:0}) = \frac{p_{\epsi}[\y_{k} - \mathcal{H}_k(\x_k)]p(\x_k\vert\y_{k-1:0})}{\int \! \mathrm{d}\x_k \, p_{\epsi}[\y_{k} - \mathcal{H}_k(\x_k)]p(\x_k\vert\y_{k-1:0})}\, ,
\end{equation}
while in the prediction step one integrates the Chapman-Kolmogorov equation \eqref{Chap-Kol}, with $l\rightarrow k$ and $k\rightarrow (k-1)$.

Finally, using  recursively the Bayes formula in the interval  $\{ t_0 \le t \le t_k\}$, the smoothing problem can be written as
\begin{equation}
\label{smoot-Bay}
p(\x_{k:0}\vert\y_{k:0}) \propto p(\x_0)\prod_{l=1}^{k}{p_{\epsi}[\y_{l} - \mathcal{H}_l(\x_l)]p_{\eeta}[\x_{l} - \mathcal{M}_l(\x_{l-1})]}.
\end{equation} 

The faithful numerical implementation of Eqs. \eqref{eq:trans}--\eqref{smoot-Bay} is impossible in realistic geophysical and other high-dimensional applications of DA because the huge size of the discrete models, ${\mathcal M}_k$, and of the observation vector, $\y_{k}$, renders the accurate representation of the relevant pdfs prohibitive. This problem is usually overcome by assuming that the error statistics are Gaussian. This assumption --- along with some form of linearization of ${\mathcal M}_k$ and of ${\mathcal H}_k$ --- allows one to fully characterize the pdfs in the equations by their first two moments only, {\it i.e.}, the means and covariances. Such a characterization results in an enormous simplification when applied to high-dimensional systems, a simplification that is in fact the basis of many successful practical DA algorithms. Some of these practical issues in applying the Bayesian perspective to DA will resurface in section~\ref{sec:DAM} in computing model evidence.   

\section{Model evidence: a contextual formulation \label{sec:evidence}}

Let us generalize the definition of an observation sequence given in section~\ref{sec:DAB} to an arbitrary time interval $k:m$, with $k\ge m$, so that
\begin{equation}
\label{Full-Obs}
\y_{k:m} = \{\y_k,\y_{k-1},...,\y_{m+1},\y_{m}\}.
\end{equation} 
In particular, for $m=-\infty$ and $k=0$, the sequence (\ref{Full-Obs}) contains all the observations from the far past up to the present time $k=0$. The likelihood of the observations given $\mathcal{M}$ can be written as
\begin{equation}
\label{lkl-obs}
p(\y_{k:} \vert\mathcal{M}) = \int \! \mathrm{d}\x \, p(\y_{k:}\vert \x, \mathcal{M})p(\x\vert\mathcal{M})
\end{equation}  
where we used the abbreviated notation 
$\y_{k:-\infty}=\y_{k:}$. 

The likelihood $p(\y_{k:}\vert\ \mathcal{M})$ is referred to as {\it model evidence} and it is often used in model selection  
\citep[{e.g.,}][and references therein]{Carson_et_al_2015}. This likelihood depends on the underlying dynamics $\mathcal{M}$, but it can be  
formulated as being dependent on any hypothesis under scrutiny. When the model dynamics is ergodic, $p(\x\vert\mathcal{M})$ is the invariant distribution on the attractor and the likelihood $p(\y_{k:})$ will capture it with an accuracy that depends on the observation model in Eq.~(\ref{obs}). The model evidence $p(\y_{k:} \vert\mathcal{M})$ is obtained by integrating over $\x$ given $\y_{k:}$ and it represents the probability that the data are actually observed under the hypothesis that the model $\mathcal{M}$ is the correct one. To simplify the notation hereafter, the explicit dependence on $\mathcal{M}$ is dropped. For instance, the marginal probability of the data is denoted by $p(\y_{k:}) = p(\y_{k:} \vert\mathcal{M})$.

The distribution $p(\x)$ in Eq.~\eqref{lkl-obs} plays the role of a prior and it thus allows one to introduce additional information about the system. The choice of the prior is usually arbitrary and one can in principle use any distribution that suits a study's specific purposes. In many practical circumstances, however, the search for a good informative prior is not straightforward, although its choice may strongly, and sometimes negatively, affect the results of the study.

The use of the climatological invariant distribution for $p(\x)$ in Eq.~(\ref{lkl-obs}) has the advantage of characterizing the system globally, but it is not very informative about its specific current conditions. This is particularly true when the underlying dynamics is out of equilibrium, when it possesses multiple stationary points, or when it is subject to large deviations.
Moreover, in the case of the large-dimensional systems used in NWP and in climate prediction, as considered in this study, a proper estimate of $p(\x)$ is complicated even further by the limitations of the computational resources. 

Finally, if the system is subject to climate change driven by a time-dependent forcing --- such as anthropogenic changes in greenhouse gas and aerosol concentrations --- its dynamics has to be described self-consistently as non-autonomous, and the mere existence of a time-independent invariant set on which $p(\x)$ is defined is questionable. 
In this case, one has to rely on the concept of a pullback or random attractor that is invariant under the system's dynamics but does depend on time, {\it i.e.} both $p$ and $\mathcal M$ in Eq.~\eqref{lkl-obs} are explicit functions of $t$ \citep[{e.g.,}][]{GCS08, Chekroun_et_al_2011, Dijkstra_2013}. This non-autonomous situation is left for later investigation.

For the purpose of a time-dependent evaluation of the model evidence in the autonomous case --- in which $\mathcal M$ only depends on time due to its non-linearity, {\it i.e.}, through its dependence on $\x$ ---  we are interested in a definition narrowed to the present moment. To this end, let us condition the observational likelihood from the present $t = t_0$ to some future time $t = t_k$, on the observation sequence up to the present, so that 
\begin{equation}
\label{faith}
p(\y_{k:}) =  p(\y_{k:1}\vert\y_{0:})p(\y_{0:}) .
\end{equation}

The assumption here %% , our {\it leap of faith}, 
is to use $p(\y_{k:1}\vert\y_{0:})$ instead %% place 
of $p(\y_{k:})$, and %% we name 
the conditional pdf $p(\y_{k:1}\vert\y_{0:})$ will be called the {\it contextual model evidence} (CME).
Implicit in Eq. (\ref{faith}) is the idea that the informational content from past observations is propagated forward by conditioning on $\y_{0:} $. While $\y_{k:1}$ is still used to diagnose the evidence in the time interval from $t_1$ to $t_k$, $\y_{0:}$ allows us now to specify the {\em context}. 

By marginalizing with respect to $\x_0$, one can write the CME as
\begin{align}
\label{cont-mod-evd}
 p(\y_{k:1}\vert\y_{0:}) & = \int \! \mathrm{d}\x_{0} \, p(\y_{k:1}\vert {\bf x}_{0},\y_{0:})p({\bf x}_{0}\vert \y_{0:})  \nn  
&= \int \! \mathrm{d}\x_{0} \, p(\y_{k:1}\vert {\bf x}_{0})p({\bf x}_{0}\vert \y_{0:}) .
\end{align}
Equation (\ref{cont-mod-evd}) shows that the CME depends on two factors, on the conditional pdf $p(\x_0\vert\y_{0:})$, which plays the role of the prior and substitutes $p(\x)$ in Eq. (\ref{lkl-obs}), and $p(\y_{k:1}\vert {\bf x}_{0})$, the likelihood of the observational sequence $\y_{k:1}$ conditioned on the system's state. 

The key point here is that the new prior, $p(\x_0\vert\y_{0:})$, is easier to compute than the invariant measure $p(\x)$. Indeed, we saw in section~\ref{sec:DAB} that the conditional pdf $ p(\x_0\vert\y_{0:})$ is the posterior density in a Bayesian inference process designed to estimate $\x_0$ based on $\y_{0:}$ and that this posterior pdf is the standard, albeit approximate, outcome of applying a DA algorithm to the model $\mathcal M$ and the data $\y_{k:1}$.
When such a forecast--assimilation cycle as described in section~\ref{sec:DAB} is routinely running --- as is the case in an operational NWP center --- an approximation of $p(\x_0\vert\y_{0:})$ is already at hand. Given the model and the observational network, the level of accuracy of the DA-based approximation for $p(\x_0\vert\y_{0:})$ is related to the degree of sophistication of the DA scheme adopted. 
We shall show in section \ref{sec:DAM} that DA can also be used to estimate the other term in Eq. \eqref{cont-mod-evd}, namely $p(\y_{k:1}\vert {\bf x}_{0})$, and thus to fully accomplish the task of estimating the CME, consistently and routinely.

\section{Data assimilation for model evidence \label{sec:DAM}}

Estimating the CME amounts to computing the integral in Eq.~(\ref{cont-mod-evd}). Analytic solutions can only be obtained for elementary cases, and the problem becomes rapidly intractable as one moves toward realistic situations with practical relevance. Numerical methods are thus necessary in practice and their degree of complexity grows with the dimension $M$ of the model and that of the data set, $d$.
For high-dimensional systems with large $M$, Monte Carlo methods using importance sampling are a viable approach, but their convergence as the sample size increases is usually very slow even when $M$ is only moderately large. 
An approximate Monte Carlo suitable for NWP applications is described in Sect.~\ref{ssec:ef}.
Large dimensionality $d$ of the data calls for Laplace method, in which the integrand is approximated as
a Gaussian and a solution for the integral can be found as a function of the mode and the covariances of this normal
distribution. We will make use of the Laplace approximation in combination with smoothers in Sect.~\ref{ssec:smoothers}.

In the geosciences, one often encounters both conditions, with $M$ and $d$ up to $O(10^9)$ and $O(10^7)$, respectively, and computing the integral in Eq.~(\ref{cont-mod-evd}) is a very challenging task that requires a trade-off between accuracy and computational efficiency. In this section, we present a hierarchy of methods based on DA practice that allow one to evaluate the CME integral in Eq.~(\ref{cont-mod-evd}) within a good approximation, and that are suitable for the large systems and big data sets typical of the environmental sciences.  To a certain
extent, the accuracy of the methods described below can be ranked according to the degree of sophistication of the DA approach on which they are based.

\subsection{General setting}
\label{ssec:setting}
We start by providing an iterative formula that will be used later, and that allows one to decompose the contextual evidence pdf as
\begin{align}
  p(\y_{K:1}|\y_{0:}) & = p(\y_{K:2}|\y_{1:})p(\y_1|\y_{0:}) \nn
  & = p(\y_{K:3}|\y_{2:})p(\y_2|\y_{1:}) p(\y_1|\y_{0:}) 
\end{align}
and so on, up to time $t_K$: 
\be
\label{eq:factor}
p(\y_{K:1}|\y_{0:}) = \prod_{k=1}^K p(\y_k|\y_{k-1:}) .
\ee
Hence the contextual evidence of the sequence $\y_{K:0}$ can be written as the product of single contextual evidences, one for each $\y_k$. Moreover, the individual contextual evidence $p(\y_k|\y_{k-1:})$ is often a tractable output of a DA scheme. 
After marginalizing over the state vector $\x_k$, we get
\be
p(\y_k|\y_{k-1:}) = \int \! \mathrm{d}\x_k \, p(\y_k|\x_k) p(\x_k|\y_{k-1:})\,,
\ee
where $p(\y_k|\x_k)$ is the observation likelihood and $p(\x_k|\y_{k-1:})$ is the forecast state pdf at $t_k$. 
\citet{Carson_et_al_2015} recently used the identity in Eq.~(\ref{eq:factor}), whereas \citet{DelMoral_2004} provides an alternative proof of its validity. 

We assume that, at any arbitrary time $t_0$, an estimate of the posterior density $p(\x_0\vert\y_{0:})$ is available as the outcome of a forecast--assimilation cycle up to $t_0$. This posterior is then used as a prior in the estimation of the CME in Eq.~(\ref{cont-mod-evd}). Unless otherwise stated, a DA method that uses first- and second-order error moments is adopted herein, so that the posterior pdf is approximated as a Gaussian, $p(\x_0\vert\y_{0:}) \approx p^{{\rm DA}}=\mathcal{N}(\x_0,\bP^\mathrm{a})$, with mean $\x_0$ and covariance matrix $\bP^{\mathrm{a}}$, where the superscript stands for analysis. Similarly, let us define $\bP^{\mathrm{f}}$ and $\bR$, to be used in the following sections, as the forecast and observation error covariance, respectively.

The {\em evidencing window} is defined as the interval $[t_0,t_K]$, and we aim at estimating the CME
\begin{equation}
\label{CME}
 p(\y_{K:1}\vert\y_{0:}) = \int \! \mathrm{d}\x_{0} \, p(\y_{K:1}\vert {\bf x}_{0})p({\bf x}_{0}\vert \y_{0:}).
\end{equation}
The forward model, Eq.~(\ref{model}), is assumed here to be deterministic and perfect, so that $\eeta_{k}=0$. The implications of this choice are discussed in section~\ref{sec:concl} and a follow-on study will relax these assumptions.   
See also \citep[][Chapter 9.1]{Reich-Cotter-2015} for a definition of model evidence in the case of a stochastic forward model. 

\subsection{Monte Carlo and importance sampling}
\label{ssec:ef}

The CME integral, Eq.~(\ref{CME}), can be estimated using a Monte Carlo approach with importance sampling. Samples are drawn from the Gaussian pdf $p^{{\rm DA}}=\mathcal{N}(\x_0,\bP^\mathrm{a})$, outcome of a Gaussian DA scheme at time $t_0$, used as proposal density for the unknown actual $p(\x_0\vert\y_{0:})$. 

Let us suppose that, at $t_0$, a sample of $N$ members, $\{\x_0^i :  i=1,\ldots, N\}$, is drawn from $p(\x_0\vert\y_{0:})$, and its members are used as initial conditions for a forward integration over the evidencing window. The CME can then be estimated as
\begin{equation}
\label{IS}
 p(\y_{K:1}\vert\y_{0:}) \approx \frac{1}{N}\sum_{i=1}^{N} p(\y_{K:1}\vert\x_{0}^i). 
\end{equation} 
The approximation in Eq.~(\ref{IS}) gets progressively better by increasing $N$, but the computational resources usually set a bound to the size of the sample. 
We will use this Monte Carlo estimate, Eq.~(\ref{IS}), with $N=10^6$, for some cases in Sect.~\ref{secc:Num-comp}. 
Equation (17), but for imperfect models, is at the basis of a model evidence computing algorithm described in \citet[][Algorithm 9.2]{Reich-Cotter-2015}. To prevent filter degeneracy and enhance efficiency with an affordable number of members/particles, they suggest the use of sequential Monte Carlo methods with resampling.

We are interested here in a straightforward use of the approximation Eq.~(\ref{IS}) in a high-dimensional context where the sample's size is unavoidably very small. 
In the geosciences, one often uses an {\em ensemble Kalman filter} \citep[EnKF,][]{evensen2009} with $N$ members to assimilate data up to the beginning $t_0$ of the evidencing window. In any realistic setup, $N\ll M$ and these $N$ members provide a reduced-order, but dynamically consistent, representation of $p(\x_0\vert\y_{0:})$ that can be used in Eq.~(\ref{IS}) in place of a random draw. 
This approach, conveniently referred here to as importance sampling (IS), albeit not very accurate, is computationally affordable in NWP and climate prediction centers. We will use it here in comparison with some Gaussian ensemble-DA methods in the numerical comparison described in Sect.~\ref{sec:results}.

We turn now to the description of several methods to compute the CME, each of which is based on a distinct Gaussian ensemble-DA algorithm.

\subsection{Filtering: Kalman filter and ensemble Kalman filter}
\label{ssec:KF}

{\bf Kalman Filter (KF).} Assuming the evolution and observation models in Eqs.~(\ref{model}) and (\ref{obs}) are both linear, and that the observation errors and the initial errors are Gaussian, the Kalman filter \citep[KF:][]{kalman1960, GMR91} is optimal. %% with these assumptions.  
In the logarithmic formulation of the KF, the observation likelihood pdfs are
\begin{align}
\ln p(\y_k|\x_k) = & -\frac{1}{2} \left\| \y_k - \bH_k\x_k \right\|^2_{\bR_k} \nn
 & - \frac{d}{2}\ln(2\pi) - \frac{1}{2}\ln \left|\bR_k\right|, \quad k=1,\ldots,K,
\end{align}
and the forecast pdfs are of the form
\begin{align}
\ln p(\x_k|\y_{k-1:}) = & -\frac{1}{2} \left\| \x_k - \x_k^\mathrm{f} \right\|^2_{\bP^\mathrm{f}_k} \nn
 & - \frac{M}{2}\ln(2\pi) - \frac{1}{2}\ln \left|\bP^\mathrm{f}_k\right|, \,\,\,  k=1,\ldots,K;
\end{align}
here $\bH_k$ is the linearized observation operator at time $t_k$. The weighted Euclidean norm $\left\| \x \right\|^2_\bA = \x^\T \bA^{-1}\x$ is used, while $\left| \bA \right|$ indicates the determinant of $\bA$.

The contextual evidence  
\be
p(\y_k|\y_{k-1:}) = \int \! \mathrm{d}\x_k p(\y_k|\x_k)p(\x_k|\y_{k-1:})
\ee 
is, in this case, the product of two Gaussian pdfs, and hence it is itself Gaussian. Besides, the two statistical moments that suffice to characterize it are given by those of the innovations: $\bH_k\x_k^\mathrm{f}$ for the mean and $\bR_k + \bH_k \bP^\mathrm{f}_k\bH_k^\T$ for the covariance matrix. As a result
\begin{align}
  \ln p(\y_k|\y_{k-1:}) =& -\frac{1}{2} \left\| \y_k - \bH_k\x_k^\mathrm{f} \right\|^2_{\bR_k+ \bH_k \bP^\mathrm{f}_k \bH_k^\T} - \frac{d}{2}\ln(2\pi) \nn
  & - \frac{1}{2}\ln \left|\bR_k+ \bH_k \bP^\mathrm{f}_k \bH_k^\T\right|, \,\,\,  k=1,\ldots,K,
\end{align}
so that the factorization formula in Eq.~(\ref{eq:factor}) now reads
\be
\label{eq:kf}
p(\y_{K:1}|\y_{0:}) = \prod_{k=1}^K \frac{\exp \( -\frac{1}{2} \left\| \y_k - \bH_k\x_k^\mathrm{f} \right\|^2_{\bR_k+ \bH_k \bP^\mathrm{f}_k \bH_k^\T} \)}{\sqrt{(2\pi)^d \left|\bR_k+ \bH_k \bP^\mathrm{f}_k \bH_k^\T\right|}}.
\ee
An alternative, less direct, proof of Eq.~(\ref{eq:kf}) can be found in \citet{Hannart-et-al-2016}.

{\bf Ensemble Kalman Filter (EnKF).}
If the forward model ${\mathcal M}(\x_k)$ in Eq.~(\ref{model}) is nonlinear, and even when both initial errors $\bP^\mathrm{f}_0$ and observation errors $\bR_k$ are Gaussian, the evolution of the forecast and analysis pdfs will not remain Gaussian, with very few exceptions. There are many ways of approximating their evolution in time, including the {\em extended KF} \citep[EKF:][]{jazwinski1970, Milleretal94}, the EnKF \citep{evensen2009} and the {\em unscented KF} \citep[UKF:][]{Grewal01}. Of these, the EKF is probably most widely used in engineering applications, while the EnKF is very widely used in the geosciences where the typical problem size is much higher.

To approximate the CME in nonlinear cases, we consider here the EnKF and introduce the matrix ${\bf E}_k = \left[\x_1,\ldots,\x_N \right]$, whose columns contain the ensemble members, as well as the corresponding normalized forecast anomalies with respect to the ensemble mean $\bX_k = {\bf E}_k\({\bf I}_N - \b1\b1^T/N\)/\sqrt{N-1}$; here $\b1\in{\mathbb R}^N$ is the column vector of ones, and ${\bf I}_N$ is the $N \times N$ identity matrix. We obtain
\be
\label{eq:enkf}
p(\y_{K:1}|\y_{0:}) \simeq \prod_{k=1}^K \frac{\exp \( -\frac{1}{2} \left\| \y_k - {\mathcal H}_k(\x_k^\mathrm{f}) \right\|^2_{\bR_k+ \bY_k\bY_k^\T} \)}{\sqrt{(2\pi)^d \left|\bR_k+ \bY_k \bY_k^\T\right|}} \, ,
\ee
where $\bY_k = \bH_k\bX_k$ is the matrix of normalized observation anomalies at $t_k$, usually estimated as $\bY_k = {\mathcal H}({\bf E}_k)\({\bf I}_N - \b1\b1^T/N\)/\sqrt{N-1}$. 

The estimate in Eq.~(\ref{eq:enkf}) is exact, and coincides with that in Eq.~(\ref{eq:kf}), when the model is linear, the initial condition and observation errors are Gaussian and when the initial ensemble anomalies span the full range of uncertainty.

\subsection{Smoothing: ensemble-4D-Var and iterative ensemble Kalman smoother}
\label{ssec:smoothers}

We will use here the Laplace approximation to estimate the CME integral Eq. (\ref{CME}). As mentioned in Sect.~\ref{sec:DAM}, in the Laplace method, the integrand is approximated as a Gaussian and the integral is a function of the mode and the covariance of this normal distribution. 
In the asymptotic limit $d\rightarrow\infty$, most of the contributions to the integral come from the vicinity of the maximum and the Laplace approximation --- whose accuracy scales as the inverse of the variance of the
approximating Gaussian --- gets progressively more accurate. 
The reader is referred to \citet{Evans_Swartz_1995} for a review of various integration methods used in statistical inference.
\citet[][Example 9.4]{Reich-Cotter-2015} suggest to use the Laplace approximation in the computation of model evidence.

In the following we describe two approaches, the ensemble 4D-Var and the iterative ensemble Kalman smoother, that can be used to compute the best estimator, the mode, and the associated uncertainty, the covariance. This latter is estimated using the Hessian of the corresponding cost-functions, in a way that is made clear later in this section.

{\bf Ensemble 4D-Var (En-4D-Var).}
%% A four-dimensional variational (4D-Var) analysis \citep{TalaCour87, GMR91} is used here with the ensemble forecast at $t_0$. The estimation and minimization processes are carried out in ensemble space, meaning that the control variable is expressed in terms of the ensemble members with corresponding ensemble coefficients, similarly to the multiple-DA (MDA) formulation of the iterative ensemble Kalman smoother (IEnKS) \citep{bocquet2013,bocquet2014,bocquet2016}. According to the nomenclature recommendations for hybrid ensemble--variational methods of \citet{Lorenc2013}, we refer to this approach as En-4D-Var.

A four-dimensional variational (4D-Var) analysis \citep{TalaCour87, GMR91} is used with background error covariances formed from the ensemble forecast at $t_0$.
According to the nomenclature recommendations for hybrid ensemble--variational methods \citep[point 7 of][]{Lorenc2013}, we refer to this approach as En-4D-Var.
Here, the estimation and minimization processes are carried out in ensemble space, meaning that the control variable is expressed in terms of the ensemble members with corresponding ensemble coefficients, similarly to the iterative ensemble Kalman smoother (IEnKS) \citep{bocquet2013,bocquet2014,bocquet2016}. However, the minimization could be performed as well with the adjoint models if available. 

In this approach, the ensemble coefficient $\w$ parameterizes the ensemble space spanned by the perturbations $\bX_0$ at $t_0$, so that $\x_0 = \barx_0 + \bX_0 \w$ at $t_0$, with $\barx_0$ being the ensemble mean at $t_0$ and, in computing $p(\y_{K:1}|\y_{0:})$, we marginalize over $\w$, 
\be
\label{eq:marginal-En-4D-Var}
p(\y_{K:1}|\y_{0:}) = \int \! \mathrm{d}\w \, p(\y_{K:1}|\w) p(\w|\y_{0:}) \, .
\ee
From the theory of the IEnKS written in ensemble space \citep{bocquet2014}, we get the first factor in the integrand of Eq.~\eqref{eq:marginal-En-4D-Var}:
\begin{align}
\ln p(\y_{K:1}|\w) = & -\frac{1}{2} \sum_{k=1}^K\left\| \y_k - {\mathcal H}_k \circ{\mathcal M}_{k:0}(\barx_0 + \bX_0 \w) \right\|_{\bR_k}^2 \nn 
 & - \frac{Kd}{2}\ln(2\pi) - \frac{1}{2}\sum_{k=1}^K\ln \left|\bR_k\right| \, , 
\end{align}
with the $\circ$ symbol representing the composition of operators.
Fixing the gauge in $\w$, {\it  i.e.} accounting for the redundant degrees of freedom in it, yields the second factor:
\be
\ln p(\w|\y_{0:}) = -\frac{1}{2} \left\| \w \right\|^2 - \frac{N}{2}\ln(2\pi) \, .
\ee
The sum of these two log-likelihoods yields the IEnKS cost function
\be\label{En4DVar_cost}
{\mathcal J}(\w) =  \ln p(\y_{K:1}|\w) + \ln p(\w|\y_{0:}) \, .
\ee

This outcome of the IEnKS variational analysis provides, without using the explicit adjoint of the model, the argument of the minimum $\w^\star$, along with the associated state vector $\x_0^\star$ and  
approximate Hessian $\bI_N + \sum_{k=1}^K\(\bY_k^\star\)^\T\bR^{-1}_k\bY_k^\star$ of the cost function ({\it i.e.} the covariance) that are required in the Laplace approximation; here $\bY_k^\star = [{\mathcal H}_k \circ {\mathcal M}_{k:0}]'_{\x_0^\star}\bX_0$, and $[{\mathcal H}_k \circ {\mathcal M}_{k:0}]'_{\x_0^\star}$ is the linearization of the nonlinear operator in the square brackets at $t_0$.
Hence, we obtain the approximation
\begin{align}
  \ln p(\y_{K:1}|\y_{0:})  \simeq & {\mathcal J}(\w^\star) + \frac{N}{2}\ln(2\pi) - \frac{1}{2}\ln \left| \partial^2 {\mathcal J}_{|\w^\star} \right| \nn
  \simeq &  -\frac{1}{2} \sum_{k=1}^K \left\| \y_k - {\mathcal H}_k \circ {\mathcal M}_{k:0}(\x_0^\star) \right\|^2_{\bR_k} \nn
         & -\frac{1}{2} \left\| \w^\star \right\|^2 - \frac{Kd}{2}\ln(2\pi) - \frac{1}{2}\sum_{k=1}^K\ln \left|\bR_k\right| \nn
         &- \frac{1}{2}\ln \left|\bI_N + \sum_{k=1}^K \(\bY_k^\star\)^\T\bR^{-1}_k\bY_k^\star\right|
 %% \, ,
\end{align}
or, exponentiating the log-likelihood,
\begin{align}
\label{eq:en-4d-var}
  &p(\y_{K:1}|\y_{0:}) \simeq \nn 
  &\frac{\exp \( -\frac{1}{2} \sum_{k=1}^K \left\| \y_k - {\mathcal H}_k \circ {\mathcal M}_{k:0}(\x_0^\star) \right\|^2_{\bR_k} -\frac{1}{2} \left\| \w^\star \right\|^2 \)}{\sqrt{(2\pi)^{Kd}\prod_{k=1}^K \left|\bR_k\right| \left|\bI_N + \sum_{k=1}^K \(\bY_k^\star\)^\T\bR^{-1}_k\bY_k^\star\right|}} \, .
\end{align}
Note that, in computing the CME with the En-4D-Var, Eq.~(\ref{eq:en-4d-var}), the initial conditions at the beginning of the evidencing window are changed at each iteration of the minimization process, and the innovations at each observation time $\{t_k :  k=1, \ldots ,K\}$ are recomputed based on the corresponding trajectory started with the new data.   

Here, the IEnKS is merely a convenient means to the solution of En-4D-Var; it is not representative of what the IEnKS has to offer, as will be seen in the following.

{\bf Iterative ensemble Kalman smoother (IEnKS).}

The IEnKS \citep{bocquet2014} allows the observations to be assimilated sequentially, one time $t_k$ after another, rather than all of them together as in the En-4D-Var.
The initial condition and the ensemble of anomalies at $t_0$ are sequentially updated by assimilating $\y_1$ in the first step,
then $\y_2$ in the second step and so on until $\y_K$. The outcomes of the analysis at step $k$ are the state $\x^\star_k$ and the normalized anomaly matrix $\bX^\star_k$, both defined at $t_0$, and both then used in the subsequent step. 
This procedure corresponds to the {\em quasi-static} IEnKS, as advocated in \citet{bocquet2014}, but it will be simply referred to as IEnKS hereafter. The IEnKS makes possible to implement Eq.~(\ref{eq:factor}), and each single contextual evidence $p(\y_k|\y_{k-1:})$ corresponds to the $k$-th analysis of the IEnKS.

At step $k$, the ensemble coefficient vector $\w_k$ is now used in the computation of $p(\y_k|\y_{k-1:})$
to parameterize the ensemble space spanned by the anomalies 
$\x_0 = \x^\star_{k-1} + \bX^\star_{k-1} \w_k$ at $t_0$, with the definition $\x_0^\star \equiv \barx_0$ and $\bX^\star_0 \equiv \bX_0$. 
Marginalizing over $\w_k$, we get
\be
\label{eq:marginal-IEnKS}
p(\y_k|\y_{k-1:}) = \int \! \mathrm{d}\w_k \, p(\y_k|\w_k) p(\w_k|\y_{k-1:}) \, .
\ee
The theory of the IEnKS written in ensemble space yields for $k=1, \ldots ,K$:
\begin{align}
\ln p(\y_k|\w_k) = & -\frac{1}{2} \left\| \y_k - {\mathcal H}_k \circ {\mathcal M}_{k:0}(\x^\star_{k-1} + \bX^\star_{k-1} \w_k) \right\|_{\bR_k}^2 \nn 
 & - \frac{d}{2}\ln(2\pi) - \frac{1}{2}\ln \left|\bR_k\right| \, ,
\end{align}
and, fixing the gauge in $\w_k$,
\be
\ln p(\w_k|\y_{k-1:}) = -\frac{1}{2} \left\| \w_k \right\|^2 - \frac{N}{2}\ln(2\pi) \, .
\ee

The sum of these two log-likelihoods yields the IEnKS cost function, as in Eq.~\eqref{En4DVar_cost}.
Then the integral in Eq.~(\ref{eq:marginal-IEnKS}) can be estimated by the Laplace method. We take advantage of the outcome of the IEnKS variational analysis to use the argument $\w_k^\star$ of the minimum and the approximate Hessian $\bI_N + \(\bY_k^\star\)^\T\bR^{-1}_k\bY_k^\star$ of the cost function required in the Laplace approximation, where $\bY_k^\star = [{\mathcal H}_k \circ{\mathcal M}_{k:0}]'_{\x_k^\star}\bX^\star_{k-1}$. 
One thus obtains the approximation: for $k=1,\ldots,K$,
\begin{align}
  \ln p(\y_k|\y_{k-1:}) \simeq & -\frac{1}{2} \left\| \y_k - {\mathcal H}_k \circ {\mathcal M}_{k:0}(\x_k^\star) \right\|^2_{\bR_k}
-\frac{1}{2} \left\|\w_k^\star\right\|^2  \nn
& - \frac{d}{2}\ln(2\pi) - \frac{1}{2}\ln \left|\bR_k\right| \nn
& - \frac{1}{2}\ln \left|\bI_N+ \(\bY_k^\star\)^\T\bR_k^{-1}\bY_k^\star\right| \, .
\end{align}
This can be re-arranged differently by using the value of $\w_k^\star$,
\begin{align}
\w_k^\star =& \left\{\bI_N+(\bY_k^\star)^\T\bR_k^{-1}\bY_k^{\star}\right\}^{-1}\bR_k^{-1}\left\{ \y_k - {\mathcal H}_k \circ {\mathcal M}_{k:0}(\x_k^\star) \right\} \nn
= &(\bY_k^\star)^\T\left\{ \bR_k+\bY^\star_k(\bY_k^\star)^\T\right\}^{-1}\left\{ \y_k - {\mathcal H}_k \circ {\mathcal M}_{k:0}(\x_k^\star)\right\}\,,
\end{align}
and gathering the first two terms to yield
\begin{align}
  \ln p(\y_k|\y_{k-1:})  = & -\frac{1}{2} \left\| \y_k - {\mathcal H}_k \circ {\mathcal M}_{k:0}(\x_k^\star) \right\|^2_{\bR_k+ \bY_k^\star\(\bY_k^\star\)^\T} \nn
  & - \frac{d}{2}\ln(2\pi) - \frac{1}{2}\ln \left|\bR_k+ \bY_k^\star\(\bY_k^\star\)^\T\right| \, .
\end{align}
%% so that 
The factorization formula in Eq.~(\ref{eq:factor}) reads therewith
\begin{align}
\label{eq:ienks}
&p(\y_{K:1}|\y_{0:}) \simeq \nn
&\prod_{k=1}^K \frac{\exp \( -\frac{1}{2} \left\| \y_k - {\mathcal H}_k \circ {\mathcal M}_{k:0}(\x_k^\star) \right\|^2_{\bR_k+ \bY_k^\star\(\bY_k^\star\)^\T} \)}{\sqrt{(2\pi)^d \left|\bR_k+ \bY_k^\star\(\bY_k^\star\)^\T\right|}} \, .
\end{align}
If the models are linear, this formula coincides with Eq.~(\ref{eq:enkf}) in the limit in  which the DA window vanishes.
Equation~(\ref{eq:ienks}) differs from Eq.~(\ref{eq:en-4d-var}) by the observations being assimilated sequentially rather than all in one batch.
%%%%%%%%%%%%

\section{Numerical results \label{sec:results}}

This section presents the numerical results of the following four approaches to computing the CME
\begin{enumerate}
\item importance sampling (IS), Eq.~(\ref{IS}); 
\item ensemble Kalman filter (EnKF), Eq.~(\ref{eq:enkf}); 
\item ensemble 4D-Var (En-4D-Var), Eq.~(\ref{eq:en-4d-var}); and
\item iterative ensemble Kalman smoother (IEnKS), Eq.~(\ref{eq:ienks}). 
\end{enumerate}

\subsection{Experimental Set-up}
\label{sec:expsetup}

{\bf Design and performance evaluation.}
We evaluate the relative performance of the four methodologies by applying them to two prototypical low-order dynamical systems often used in theoretical NWP and climate studies. 
Given the potential uses of CME mentioned in section~\ref{sec:intro}, we wish to evaluate the accuracy and robustness of the proposed schemes when observations are consistent with the model ({\it i.e.} perfect model), 
but also when they are not, insofar as CME is intended for the purpose of model evaluation and selection. It is, therefore, particularly relevant to estimate CME when the observations are not consistent with the model. 

The contextual model evidence will thus be systematically computed for two models: the {\it correct} and the {\it incorrect} one. 
To be consistent with the intended applications to detection and attribution \citep[and references therein]{Hannart-et-al-2016}, we use the notation $\mathcal{M}_1$ for the true model and $\mathcal{M}_0$ for the other one. This notation agrees with the use, in causality theory \citep{Pearl00}, of index `1' for the {\em factual} world and `0' for the {\em counterfactual} one. 
In both cases, the CME is estimated based on the same sequence of observations, which by definition are generated by the true model. Statistically speaking, the CME of observations is expected to be higher when using the true model than when using another one, as will be discussed further on. 

In order to decide which of the four approximations performs best in each of the situations being tested, we have to compare their outcome with that of alternative computational methods that are able to evaluate the CME with high accuracy. We have therefore used a massive Monte Carlo (MC, Eq. (\ref{IS})) integration with  $10^6$ particles and a high-degree integration method, the Gauss-Hermite quadrature \citep[GHQ, e.g.,][]{Liu_Pierce_1994}. The results are reported in subsection "Performance of the four methods" of Sect.~\ref{secc:Num-comp}.

Gauss-Hermite quadrature is a suitable option in the present context given that the kernel distribution, which is the posterior pdf of the underlying EnKF running with the true model (see ``Models and implementation`` subsection below), is assumed Gaussian although the unknown actual posterior distribution may well not be so.
An analytic approximation of the integral can be obtained as a function of the roots of the corresponding Gauss-Hermite polynomials (see Appendix for details). 
Both the accuracy of the method and the associated computational cost increase with the degree of the Gauss-Hermite polynomials. 
Similarly to the kernel for GHQ, the ensemble-based Gaussian posterior pdf provided by the EnKF at the beginning of the evidencing window is used as the importance density from which the $10^6$ members of the MC are randomly drawn.
%% In order to keep the numerical cost reasonable, GHQ was used here only for L63 but with polynomial degrees as high as $32$, whereas Monte Carlo integration was applied to both models. 

%% \subsubsection{Models and implementation}
{\bf Models and implementation.} The experimental setup is chosen to mimic the situation encountered in an operational NWP or climate prediction center, in which DA is routinely carried out to update the model state based on observations. We suppose that an EnKF is used for this purpose and that it assimilates noisy observations whose error is assumed to be an unbiased random Gaussian noise, so that $\epsi_{k}$ is sampled from $\mathcal{N}(\bzero,\bR_k)$.  

We adopt a standard identical-twin experiment 
configuration \citep{Bengtsson81}, in which a solution of the correct model $\mathcal{M}_1$ is taken to represent the truth and is sampled to generate synthetic observations.
The control trajectory, into which the observations are assimilated, is generated by randomly perturbing the initial condition of the true trajectory. This control trajectory then evolves using the correct model, and the synthetic observations are assimilated into it using the EnKF with the setup described below. 

\begin{figure}
\centering
\includegraphics[width=19pc]{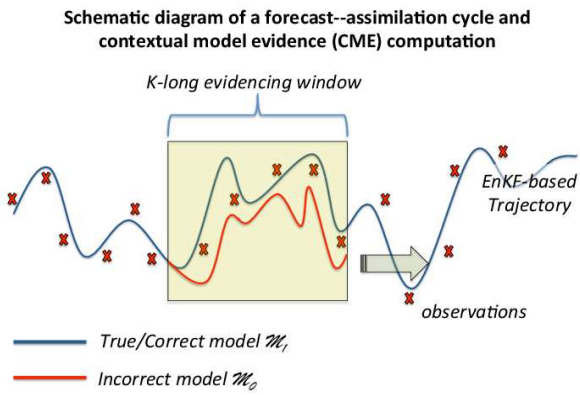}
\caption{Schematic diagram of a routine forecast--assimilation cycle that includes contextual model evidence (CME) computation run alongside the EnKF-based trajectory of the correct model $\mathcal{M}_1$ (blue) and the incorrect one $\mathcal{M}_0$ (red). }
\label{FIG1}
\end{figure}

In parallel, we carry out a sequence of model evidence diagnosis, at each observation time $t_k$ and over an evidencing window comprising $K$ observation vectors. The CME is then computed for the correct and incorrect models, $\mathcal{M}_1$ and $\mathcal{M}_0$, using the four methods. A schematic illustration of the experimental setup is given in Fig.~\ref{FIG1}. 

%%%%%%%%%%%%%%%%%%%%%%%%%%%%%%%%%%%%%%%%%%%%%%%%%%%%%%%%%%%%%%%%%%%%%%%%
%%%%%%%%%%%%%%%%%%%%%%%%%%%%%%%%%%%%%%%%%%%%%%%%%%%%%%%%%%%%%%%%%%%%%%%%

Experiments are conducted using two low-order nonlinear chaotic models widely used in the predictability and DA literature: (i) the Lorenz 3-variable convection model \citep[L63:][]{Lorenz1963}; and (ii) the Lorenz 40-variable mid-latitude atmospheric dynamics model \citep[L95:][]{lorenz1998}.

The original L63 model is modified to include an additional time-constant forcing, cf. \cite{Palmer1999}, as follows:
\begin{align}  
\label{L63}
 \frac{{\rm d}x}{{\rm d}t} & = \sigma(y-x)+\lambda_i\cos\theta \,, \\
 \frac{{\rm d}y}{{\rm d}t} & = \rho x-y-xz+\lambda_i\sin\theta \,, \qquad i=0,1 \\
 \frac{{\rm d}z}{{\rm d}t} & = xy - \beta z \, .
\end{align}  
The canonical values $(\sigma, \rho, \beta)=(10, 28, 8/3)$ for the standard coefficients are used, while the parameter $\lambda$ modulates the strength of the external forcing; it is chosen to be $\lambda_1=0$ for the correct model, and $-8 \le \lambda_0 \neq 0 \le 8$ for the incorrect one. The coefficient $\theta$ gives the angle of the forcing and is set to $\theta=7\pi/9$ as in \cite{Palmer1999}.

The equations of L95 read \citep{lorenz1998}: for $j=1,\ldots ,M$,
\begin{align}
\frac{{\rm d}x_j}{{\rm d}t} & = x_{j-1}\left(x_{j+1} -x_{j-2}\right)-x_j+F_i \,,  \qquad i=0,1 
\end{align}
with $M=40$ and periodic boundary conditions, $x_0=x_M$, $x_{-1}=x_{M-1}$ and $x_{M+1}=x_{1}$. The standard value $F_1 = 8$ is used for the correct configuration, while $5 \le F_0 \neq 8 \le 11$ for the incorrect one. For both models, L63 and L95, the range of values chosen for the forcing in the incorrect configuration is such that the
overall asymptotic stability properties of the system do not differ substantially from the true configuration. In all the situations we have examined, the true and perturbed models are both chaotic, although with a different spectrum of Lyapunov exponents.

The EnKF is used to assimilate the observations into the basic trajectory of the true model, shown in blue in Fig.~\ref{FIG1}. We use an 
ensemble square-root Kalman filter implementation, the ensemble transform Kalman filter \citep[ETKF:][]{hunt2007}. The numerical
setup for the models and the EnKF is as follows.
\begin{itemize}
%% \item Numerical integration scheme: fourth-order Runge-Kutta with time step $\delta t = 0.01$ for L63 and $\delta t = 0.05$ for L95.
\item Number of ensemble members: $N=4$ for L63; $N=20$ for L95.
\item Observation distribution: both models are fully observed, {\it i.e.} $d=3$ for L63 and $d=40$ for L95.
  The time interval between updated is $t_{k+1}-t_k=0.10$ for L63 and $t_{k+1}-t_k=0.05$ for L95. 
\item Observation error: unbiased Gaussian white noise with covariance $\bR = \sigma_\mathrm{obs}^2{\bf I}_d$ with $\sigma_\mathrm{obs}=2$ for L63, and $\sigma_\mathrm{obs}=1$ for L95.
\item Forecast error covariance inflation factor: $\bP^\mathrm{f} \rightarrow \alpha^2 \bP^\mathrm{f}$, with $\alpha=1.03$ for both models.
\end{itemize}

The experiments are ran after a 2,000 time step--long spin-up that is not taken into account in
computing the statistics. The CME values are computed over $K$-long evidencing windows, starting at each observation
time $t_k$, and using the four methods under comparison: IS, EnKF, En-4D-Var and IEnKS, for both the ``factual'' scenario, with $\mathcal{M}_1$, and the ``counterfactual'' one, with $\mathcal{M}_0$ (see Fig.~\ref{FIG1}). 
In both cases, the same set of observations, sampled from the
underlying true evolution, are used. A consequence of this latter assumption is that the correct model CME is generally
larger than the incorrect one: observations are more likely to belong to the true rather than to the perturbed
world. In the following, we will compute and show the logarithm of the CME, {\it i.e.} the logarithm of Eq.~(\ref{CME}), but will indistinguishably refer to it as CME.

\subsection{Comparing the methods and CME values}
\label{secc:Num-comp}

{\bf Performance of the four methods.}
In this subsection, we are interested in the accuracy of the four methods. To this end, they are compared here with MC and GHQ (see Sect.~\ref{sec:expsetup}). 
The experiments last 200 time steps (after the 2,000 long spin-up), the evidencing
window is $K=10$, and the CMEs are computed at each time step. The forcing strength in the incorrect model is $\lambda_0=8$ and $F_0=11$ for L63 and L95,
respectively. 

To reduce the computational burden, GHQ is used only for L63  but with polynomial degrees as high as $32$, while MC is used for both L63 and L95. MC and GHQ will not necessarily return the same estimate of the actual CME and, when necessary, we will assume GHQ to be the most accurate. To clarify this issue, we have computed the MC estimates as a function of $N$ in the range $[10^2,10^6]$, along with the best-fit to a power law of the form, $y(x)=a+bx^c$. This allows to extrapolate the  asymptotic limiting value for $N\to\infty$, MC$^{\infty}$, by taking the limit $y^{\infty}=\lim_{x\to\infty}y(x)$. The results of this analysis are reported in Table 1. The root-mean-square-error (RMSE) of the best-fit is $O(10^{-2})$ for the correct L63  and $O(10^{-1})$ for both correct and incorrect L95. The incorrect L63 model is unsurprisingly the most intricate to treat and the RMSE is $O(10^{0})$. In all cases the best-fit coefficient $c$ is negative, so that the asymptotic values, MC$^{\infty}$, are retrieved from the best-fit coefficient $a$. The extrapolated asymptotic values can be compared with the estimates given by GHQ and MC (with $N=10^6$) for L63, and with MC ($N=10^6$) for L95. We see that for the L63 correct case, MC$^{\infty}$, MC (N=10$^6$) and GHQ all coincide and are thus indistinguishably good estimates of the target CME value to compare with the DA-based estimates. 
The numerical results for L63 incorrect model suggest that convergence has not yet been reached, as it is also reflected by the lower accuracy of its best-fit. 
The estimated limiting value, MC$^{\infty}$, is slightly closer to GHQ than MC with $N=10^6$, and GHQ is taken as the reference here.
For the L95, both correct and incorrect models, the asymptotic estimates, MC$^{\infty}$, are very close to those for $N=10^6$, particularly for the correct model case as expected. We are thus confident that, in this case, MC with $N=10^6$ provides reasonably good estimates of the actual CME to compare with the DA-based computations.

\begin{table*}
\caption{Average correct ($\log p_1$) and incorrect ($\log p_0$) CME estimated using Gaussian Hermite Quadrature (GHQ) with degree $32$ (see text and Appendix for details) and Monte Carlo with $10^6$ particles (MC). The RMSE of the best-fit to the power law, $y(x)=a+bx^c$, for the numerical values of MC with $10^2\le N\le10^6$ are given in the $5^{{\rm th}}$ column. The extrapolated asymptotic value for $N\to\infty$ is given in the $6^{{\rm th}}$ column.}
\centering
\begin{tabular}{|M{0.8cm}|M{4cm}|M{1.5cm}|M{2cm}|M{3cm}|M{1.5cm}|}
\toprule
  &  & GHQ & MC (N=10$^6$) &  RMSE of best-fit to $y=a+bx^c$ & $\mathrm{MC}^{\infty}$ \\
\midrule
         & Correct Model $\mathcal{M}_1$  & -65.44  & -65.44 & 0.004 & -65.44 \\
L63 & & & & & \\
         & Incorrect Model $\mathcal{M}_0$ & -78.19 & -109.19 & 2.22 &-102.25 \\
\midrule
         & Correct Model $\mathcal{M}_1$  & -- & -574.57 & 0.411 & -574.63    \\
L95 & & & & & \\
         & Incorrect Model $\mathcal{M}_0$ & -- & -744.68 & 0.440 & -729.25  \\
\bottomrule
\end{tabular}
\label{table1}
\end{table*}

%% \begin{table*}
%% \caption{{\ac Average correct ($\log p_1$) and incorrect ($\log p_0$) CME estimated using Gaussian Hermite Quadrature (GHQ) with degree $32$ (see text and Appendix for details) and Monte Carlo with $10^6$ particles (MC). The best-fit coefficients to the power law, $y(x)=a+bx^c$, for the numerical values of MC with $10^2\le N\le10^6$ are given in the $5^{{\rm th}}$ column, along with the root-mean-square-error of the fit. The extrapolated asymptotic value for $N\to\infty$ is given in the $6^{{\rm th}}$ column.}}
%% \centering
%% \begin{tabular}{|M{0.8cm}|M{4cm}|M{1.5cm}|M{2cm}|M{3cm}|M{1.5cm}|}
%% \toprule
%%   &  & \bf{GHQ} & \bf{MC} (N=10$^6$) & {\bf Best-fit to} ${\bf y=a+bx^c}$ & $\bf{MC^{\infty}}$ \\
%% \midrule
%%          & \textbf{Correct Model} $\mathcal{M}_1$  & -65.44  & -65.44 & a=-65.44 $\quad$  b=-1.23 c=-0.52 RMSE=0.004 & -65.44 \\
%% \bf{L63} & & & & & \\
%%          & \textbf{Incorrect Model} $\mathcal{M}_0$ & -78.19 & -109.19 & a=-102.25 b=-179.80 c=-0.25 RMSE=2.22 &-102.25 \\
%% \midrule
%%          & \textbf{Correct Model} $\mathcal{M}_1$  & -- & -574.57 & a=-574.63 b=-35.04 c=-0.31 RMSE=0.411 & -574.63    \\
%% \bf{L95} & & & & & \\
%%          & \textbf{Incorrect Model} $\mathcal{M}_0$ & -- & -744.68 & a=-729.25 b=-114.63 c=-0.14 RMSE=0.440 & -729.25  \\
%% \bottomrule
%% \end{tabular}
%% \label{table1}
%% \end{table*}

\begin{figure}
\centering
\includegraphics[height=16pc,width=21pc]{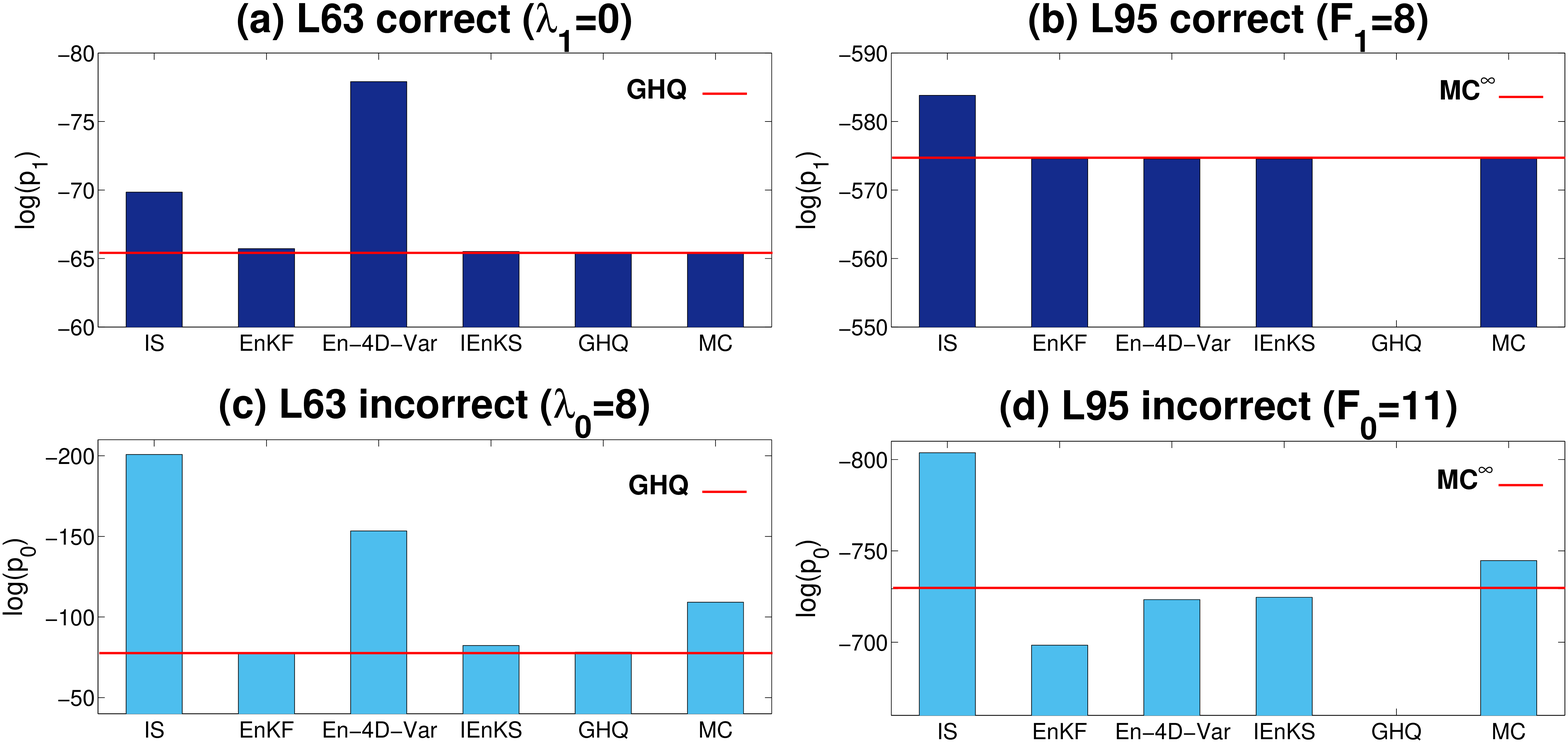}
\caption{Average CME value for the correct and incorrect model, {\it i.e.} $\log(p_1)$ vs. $\log(p_0)$, estimated using our four methods --- IS, EnKF, En-4D-Var and IEnKS --- as well as the two high-accuracy validation methods, Monte Carlo (MC) with $10^6$ particles, and Gauss-Hermite quadrature (GHQ) with degree $32$. 
The horizontal line depicts the reference value given by the GHQ or MC$^{\infty}$ (see Table 1).}
\label{FIG2}
\end{figure}

Results for the mean CME using the four methods along with GHQ and MC ($N=10^6$) are displayed in Fig.~\ref{FIG2}; the red horizontal line depicts the target actual CME value. 

%%%%%%%%%%%%%%%%%%%%%%%%%%%%%%%%%%%%%%%%%%%%%%%%%%%%%%%%%%%%%%%%%%%%%%%%

%%%%%%%%%%%%%%%%%%%%%%%%%%%%%%%%%%%%%%%%%%%%%%%%%%%%%%%%%%%%%%%%%%%%%%%%

For the L63 correct model, the EnKF and IEnKS are remarkably providing almost the same estimate, the closest to the GHQ target, followed by IS. In contrast, En-4D-Var is the farthest from the target, a result that is possibly related to the effect of a non-quadratic cost function and the resulting presence of multiple minima that makes difficult the convergence to the global minimum of the cost function. The problem experienced by the En-4D-Var smoother seems to be successfully overcome by the IEnKS, by virtue of the quasi-static formulation adopted in this study \citep[see][for a definition and discussion on the quasi-static approach]{pires1996,bocquet2014}.  

MC is taken as the reference for the L95 correct case, and all four methods being evaluated, except IS, converge to the target with a similar high level of accuracy. When observations are dense and frequent enough, and when the model forecast is only weakly nonlinear in between two observation times, the errors are nearly Gaussian and the three DA methods track the true signal successfully; as a result, their CME estimates almost coincide.

Estimating the CME for the incorrect models is more intricate and, in fact, MC and GHQ no longer provide identical results for the L63 model (see also Table \ref{table1}). As explained above, GHQ is used here as the reference target value.
In contrast to the true-model case, IS is now the least accurate, EnKF and IEnKS are the best, while En-4D-Var is in between the two.

When comparing to the incorrect-model CME for the L95 model, we see that the En-4D-Var and IEnKS methods provide similarly accurate results, followed by the
EnKF, while IS is not able to converge to the target given here by the MC.  Together, the results for the true- and 
the incorrect-model CME in the L95 model suggest that, at least in this weakly nonlinear regime, the accuracy of the DA-based
estimates of the CME is connected to the level of sophistication of the DA method, with the two smoothers, En-4D-Var and IEnKS, performing better than the filter (EnKF).

\begin{figure}
\centering
\includegraphics[height=16pc,width=21pc]{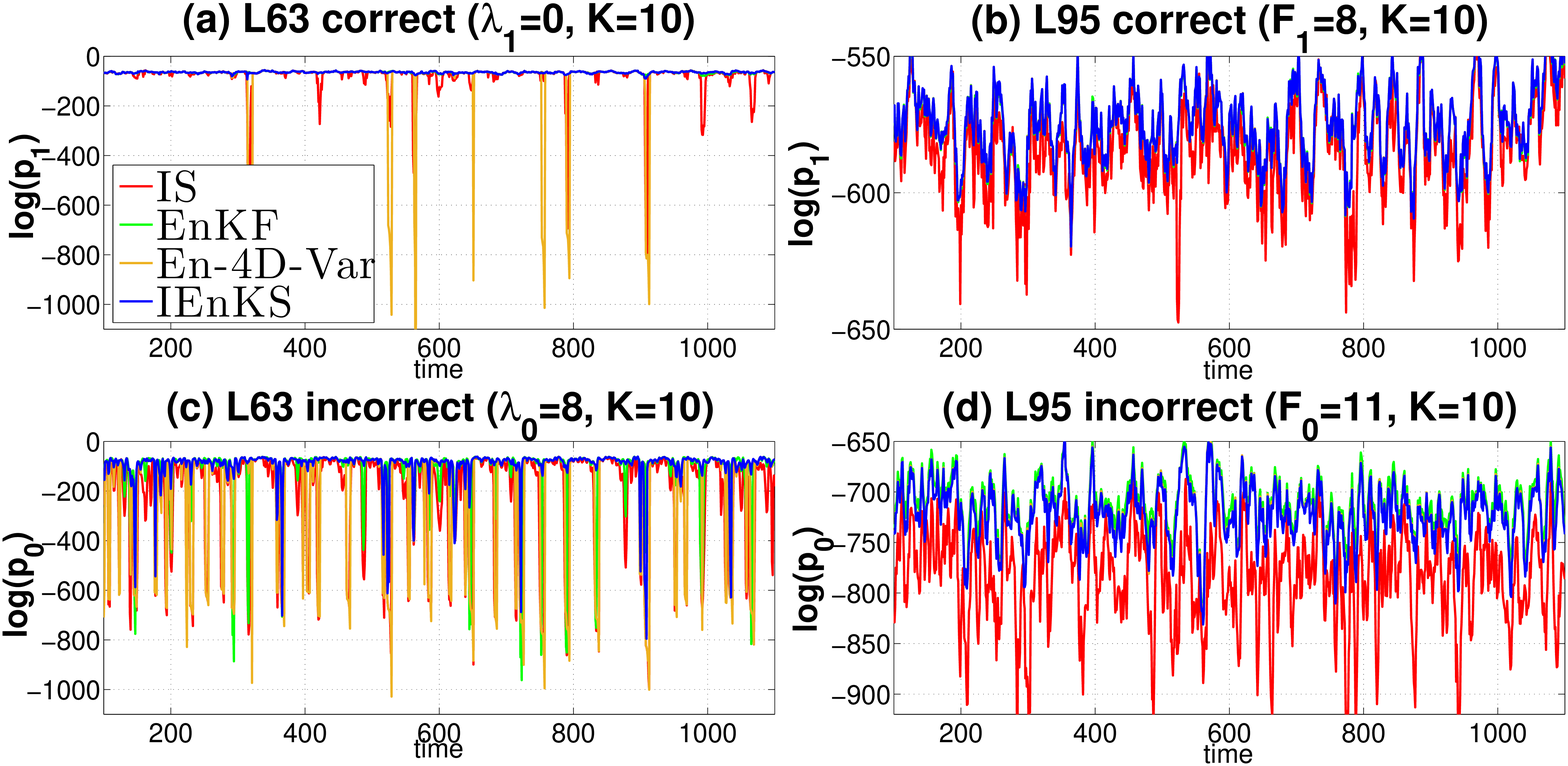}
\caption{Time series of $1000$ instantaneous CME values starting from $t=100$, for the correct (top panels) and incorrect (bottom panels) configurations of the L63 model (left panels) and L95 model (right panels), respectively; the evidencing window has length $K=10$ for all four panels. The values of the forcing in the correct vs. the incorrect models are as in Fig.~\ref{FIG2}; see also panel legends. The CME is computed using the four methods: IS, EnKF, En-4D-Var and IEnKS.}
\label{FIG3}
\end{figure}

{\bf Time series of CME values.}
Figure~\ref{FIG3} shows 1000-time steps long time series of the instantaneous CME values, for the correct and incorrect models, over the time interval 100--1100 (after the spin up) for the
L63 model (left panels) and the L95 one (right panels). The parameter values appear in the figure caption, and the evidencing window is $K=10$ in all experiments.

%%%%%%%%%%%%%%%%%%%%%%%%%%%%%%%%%%%%%%%%%%%%%%%%%%%%%%%%%%%%%%%%%%%%%%%%

%%%%%%%%%%%%%%%%%%%%%%%%%%%%%%%%%%%%%%%%%%%%%%%%%%%%%%%%%%%%%%%%%%%%%%%%

As expected, the CME values in the incorrect model are algebraically smaller than in the correct one, for all four methods, and especially so in the L95 model; see Fig.~\ref{FIG3}(d). The behavior of the CME time series is, moreover, very different between the L63 and L95 model. 
In the L63 model, the CME values are predominantly quite small in absolute value, but large on-off bursts are observed, in which the CME jumps to algebraically very low values, {\it i.e.} to quite large absolute values. These spikes are also observed in the correct model, but they get larger and more frequent in the incorrect one. Moreover, these spikes are mainly found in the IS and En-4D-Var methods, with only some occasional instances in the EnKF, and to an even lesser extent in the IEnKS. 

The CME time series of the correct L95 model in panel (b) are almost never as small in absolute value as for the L63 model in panel (a), while three of the four
methods to compute CME provide similar results, and IS alone is both negatively biased overall and exhibits strong negative spikes as well.  
The CME values for the incorrect L95 model in panel (d) are shifted downward by about 100 logits with respect to the correct model for the estimates provided by the EnKF, En-4D-Var and IEnKS methods, while the IS departures are even more substantial.

The differences in the numerical results for CME estimation with the L63 and L95 models using DA-based methods can be interpreted by the two models' dynamical and statistical properties. The L63 model has a strange attractor which is bimodal. As shown already by \cite{Milleretal94} in comparing an EKF with a 4D-Var method, these features can easily lead to situations in which the model solution and the observations are on a different ``wing of the butterfly,'' {\it i.e.} on a different lobe of the attractor, thus making already the state estimation particularly challenging. 

Such situations are even more deleterious when using DA to evaluate model evidence and when the incorrect model is under study. \citet{CVN2008} and \cite{CV2010} provide further examples of applying DA to both the L63 and L95 model, and discuss some of the implications of their dynamical features on DA results. The fine, onion-skin structure of the L63 attractor is most likely to be responsible for the large spikes in the CME values obtained by the En-4D-Var method: multiple minima in the cost function arise, and their number increases as the window length $K$ increases, cf.~Figs.~6a--c in \citet[][and discussion therein]{Milleretal94}. This inference is confirmed by some of the numerical experiments below; see Fig.~\ref{FIG6} and related discussion.  

The use of the quasi-static approximation in the IEnKS, in which one observation vector is sequentially added at each step of the minimization, helps track the global minimum of the cost function and prevents getting trapped in a secondary minimum that would give a strongly biased estimate of the CME. This may explain the relatively smoother profile of the IEnKS-based CME estimates. 

%% \subsubsection{Distribution of CME values}

\begin{figure}
\centering
\includegraphics[height=16pc,width=21pc]{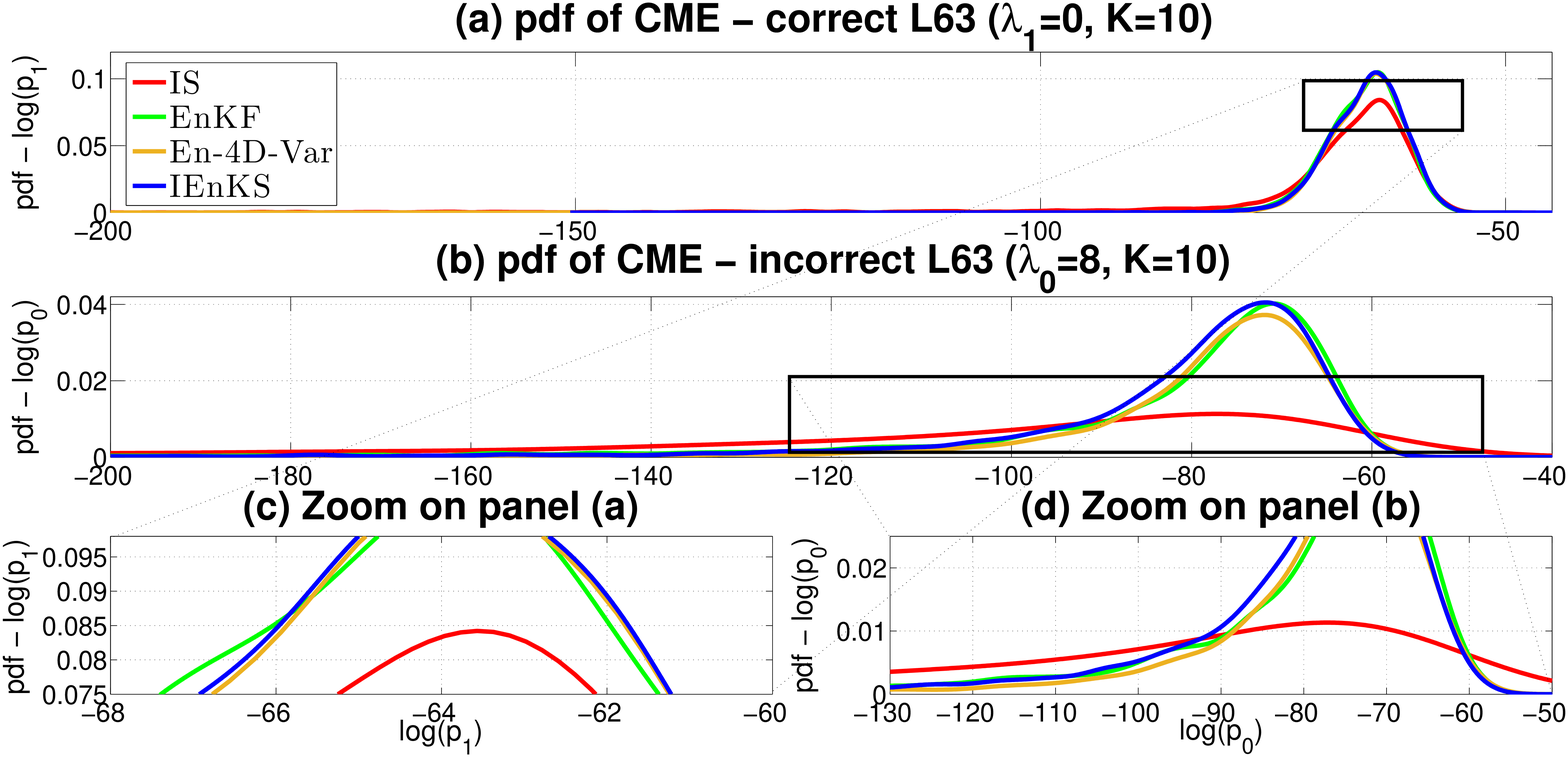}
\caption{Same as Fig.~\ref{FIG2} for the probability density function (pdf) of the CME for the L63 model, and the four methods --- IS, EnKF, IEnKS and En-4D-Var --- only. (a,b) Full pdf; and (c,d) zoom on the central part of the pdf.}
\label{FIG4}
\end{figure}

\begin{figure}
\centering
\includegraphics[height=16pc,width=21pc]{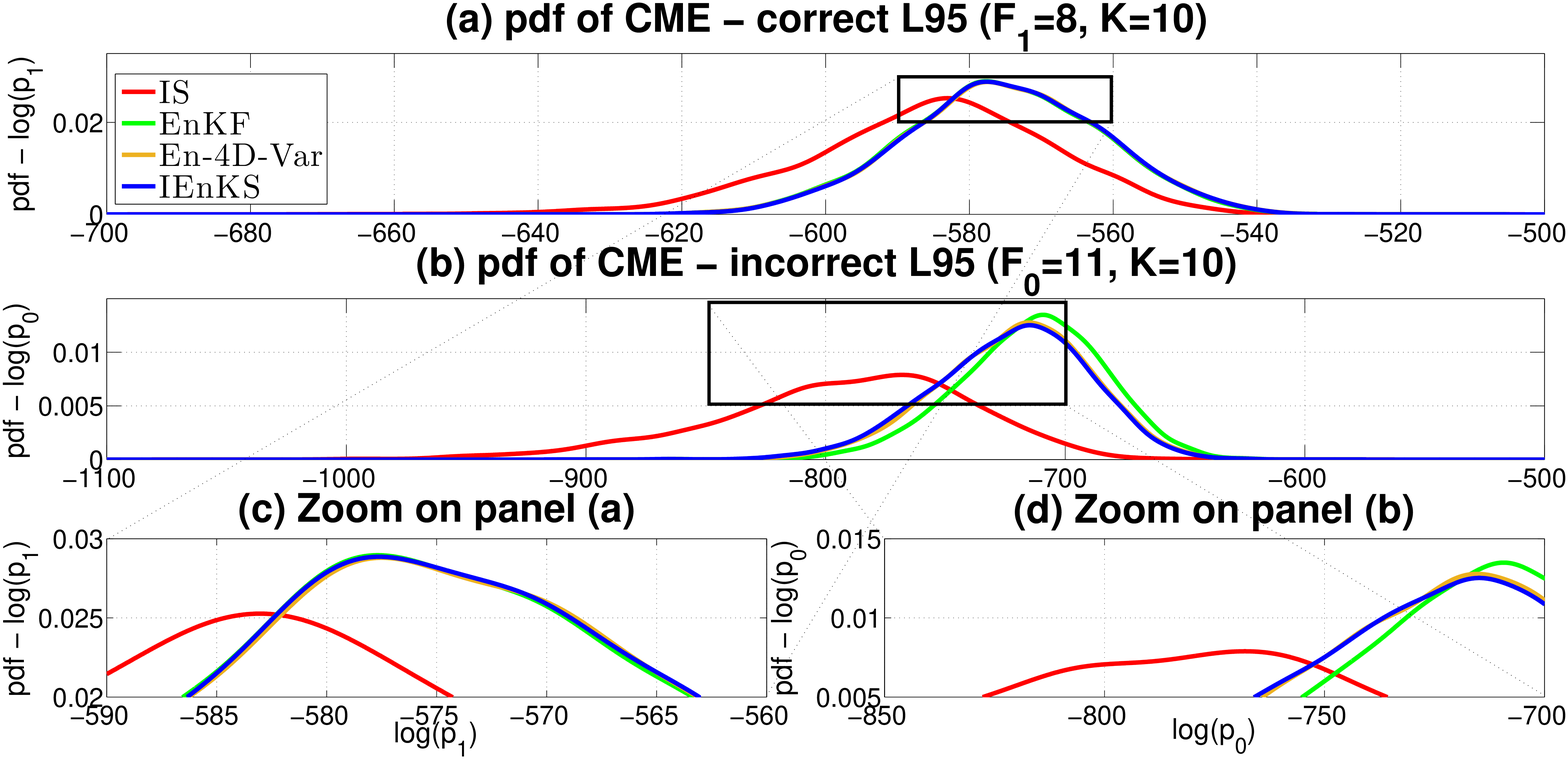}
\caption{Same as Fig.~\ref{FIG4} but for the L95 model.}
\label{FIG5}
\end{figure}

\begin{figure}
\centering
\includegraphics[height=16pc,width=21pc]{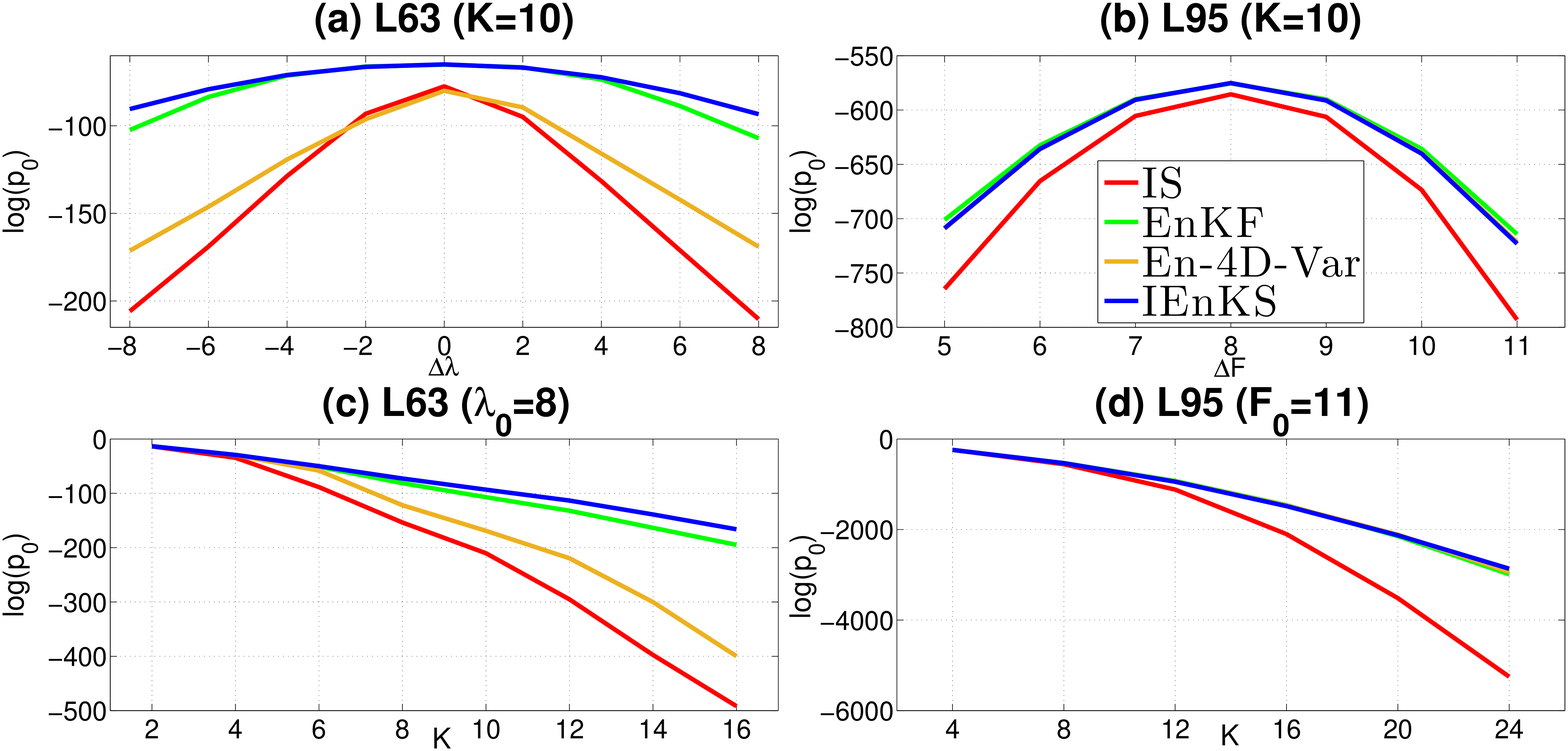}
\caption{Mean CME as a function (a,b) of the difference between the incorrect and correct model forcing (a) $\Delta\lambda=\lambda_0-\lambda_1$ and (b) $\Delta F=F_0-F_1$, and (c,d) of the length of the evidencing window. Results for (a,c) the L63 model and (b,d) the L95 model. In the top panels (a,b) the evidencing window is $K=10$, while in the bottom panels the forcing parameter equals the incorrect value: (c) $\lambda_0=8$ for L63, and (d) $F_0=11$ for L95.} 
\label{FIG6}
\end{figure}

{\bf Distribution of CME values.} The distributions of the CME values are shown in Figs.~\ref{FIG4} and \ref{FIG5} for the L63 and L95 models, respectively. The statistics are computed based on 4,000 time steps long experiments, started after the spin-up.  
The same correct and incorrect forcing values as before were used in panels (a) and (b), respectively, while the corresponding zoomed areas of the pdfs are plotted in panels (c) and (d), respectively. Note that to improve visualization of the main features of the pdfs we have intentionally limited the displayed range for the L63 to $[-200,-40]$. As a consequence the pdfs in Fig.~\ref{FIG4} do not reflect the large, occasional, peaks observed in Fig.~\ref{FIG3}a,c.

%%%%%%%%%%%%%%%%%%%%%%%%%%%%%%%%%%%%%%%%%%%%%%%%%%%%%%%%%%%%%%%%%%%%%%%%
%%%%%%%%%%%%%%%%%%%%%%%%%%%%%%%%%%%%%%%%%%%%%%%%%%%%%%%%%%%%%%%%%%%%%%%%

%%%%%%%%%%%%%%%%%%%%%%%%%%%%%%%%%%%%%%%%%%%%%%%%%%%%%%%%%%%%%%%%%%%%%%%%
%%%%%%%%%%%%%%%%%%%%%%%%%%%%%%%%%%%%%%%%%%%%%%%%%%%%%%%%%%%%%%%%%%%%%%%%

In agreement with Fig.~\ref{FIG3}, the three Gaussian DA-based CME estimates are quite similar to each other in the factual case and differ from the IS estimate, which shows a negative bias in both models, particularly in L95, cf Fig.~\ref{FIG5}(a). A closer inspection of the zoomed area for the L63 model in Fig.~\ref{FIG4}(c) reveals that the EnKF pdf is slightly shifted toward smaller values, whereas the En-4D-Var and IEnKS estimates are still very close to each other. The similarities among the three DA-based estimates are visible when looking at the zoomed area for the L95 model in Fig.~\ref{FIG5}(c).  

Differences among IS, EnKF, En-4D-Var and IEnKS estimates are, however, apparent in the CME estimates with the incorrect model. All four distributions are now displaced toward smaller values than in the corresponding correct model cases --- since the observations are more likely in the true world --- and other discrepancies appear, too. It is thus remarkable that the IS pdfs, for both the L63 and L95 models, show now even stronger departures from the other three distributions. The IS method here does not use any resampling to keep the members on the track of the true signal; this can lead to unrealistic underestimations of the CME, as noted already in Fig.~\ref{FIG3}.  

When looking more attentively at the pdfs for the three DA-based approaches, additional details emerge. In the L63 model, the IEnKS estimates differ from both the EnKF and En-4D-Var, in particular its pdf is slightly larger toward smaller CME values, cf.~Fig.~\ref{FIG4}(d), although the modes of the three distributions are almost indistinguishable, cf.~Fig.~\ref{FIG4}(b). The situation is somewhat different in the L95 model, where the En-4D-Var and the IEnKS pdfs are very close when using the incorrect model, while the EnKF pdf peaks at a slightly larger CME value.

{\bf Sensitivity analysis of the results.} 
This section describes the numerical sensitivity analysis of the CME computational methods with respect to the discrepancy between correct and incorrect forcing, $\Delta\lambda$ or $\Delta F$, and the length of the evidencing window, $K$. We wish to understand how the different methods respond to the effect that these factors have on the CME integral. The comparison among the DA-based methods is guided by the following predictions. First, the CME gets smaller for increasing $\Delta\lambda$ or $\Delta F$. In fact, if $\Delta\lambda$ or $\Delta F$ is seen as a measure of the difference between the correct and incorrect models, the marginal likelihood of observations of the latter model will decrease along with the difference between the two. Recall that "correct model" is intended here as the one used to generate the observations so that, by construction, the CME values must be higher in this case. The sensitivity analysis to $\Delta\lambda$ or $\Delta F$ aims to assess the extent to which this is the case in the DA-based approximations of the CME and how this compares among the four approaches. Second, the CME gets smaller for increasing $K$. In fact, the longer the evidencing window the longer the time at disposal for the model to manifest its deviations from the observations, and the more statistically reliable the CME should be. We do not have in this case a guess for the unknown actual growth rate, but we expect that the better the data assimilation is able to keep the incorrect model close to the observations, the bigger the CME values.

The way the four methods respond to the amplitude of the discrepancy between the correct and the incorrect forcing, and to the length $K$ of the evidencing window, is depicted in Fig.~\ref{FIG6}. 

%%%%%%%%%%%%%%%%%%%%%%%%%%%%%%%%%%%%%%%%%%%%%%%%%%%%%%%%%%%%%%%%%%%%%%%%
%%%%%%%%%%%%%%%%%%%%%%%%%%%%%%%%%%%%%%%%%%%%%%%%%%%%%%%%%%%%%%%%%%%%%%%%

In panels (a) and (b), the evidencing window is kept fixed at $K=10$, while the difference in the forcing is plotted on the abscissa. 
It is striking that, roughly speaking, the CME values for all four methods are inversely proportional to the difference, in absolute value, between the two forcings (see Fig.~\ref{FIG6}(a,b)). %Although we do not expect such a straightforward proportionality to be still the case in a more realistic model and observational scenario, it does facilitate the interpretation of the results in the present idealized setting.
%This behavior is displayed in Figs.~\ref{FIG6}(a,b). For all four estimation methods, the CME gets smaller as the difference between the correct and incorrect models increases. 
Nevertheless, the rate of change of the CME values varies from one method to another, and the differences between methods are noticeably larger in the case of the L63 model in panel (a). 
Of the four, it is IS that falls off most strongly as the forcing value deviates more from the correct one, in both panels. The IS method is thus the least accurate at a given forcing, but also the most sensitive one to the deviation of the forcing from the correct value.

The En-4D-Var method also exhibits a marked sensitivity to forcing value, although less so than the IS method. Interestingly, the En-4D-Var estimate of CME in the true case, {\it i.e.} for $\lambda_1=0$, is very close to IS, and increasingly closer to EnKF and IEnKS than IS as we move away from the correct forcing value for the L63 model, cf.~panel (a). The EnKF and IEnKS estimates peak at a slightly higher value for the true CME and show lesser dependence on the forcing difference between the two models. In the L95 model, cf.~panel (b), the three DA-based computational techniques --- EnKF, En-4D-Var and IEnKS --- give a similar response to the difference in forcing, while IS is negatively biased with respect to all three and shows a larger dependency on the forcing difference between the correct and incorrect models. 

The impact of the evidencing window length, $K$, is analyzed in panels (c) and (d), for the L63 model and the L95 one, respectively. Here the incorrect model forcing is kept fixed at $\lambda_0=8$ for L63 and at $F_0=11$ for L95. The increase of the evidencing window length facilitates the discrimination between the correct and the incorrect models, since the two dynamics have a longer time interval each to manifest their differences, but it also makes the computation of the integral in Eq.~\eqref{CME} more delicate, calling therewith for more sophisticated integration methods. 

The L63 model highlights this quandary better, due to the model's greater simplicity but also bimodal character, cf.~panel (c). In it, we see how IS changes rapidly by increasing $K$, but En-4D-Var is also very sensitive. While the correctness of the En-4D-Var estimate cannot be assessed, the large deviation from the other two DA-based methods points to its lower accuracy.
%% As suggested by the results of \cite{Milleretal94} and here in connection with Fig.~\ref{FIG3},
Again, the gap between the En-4D-Var estimates and those of EnKF and IEnKS may well be due to the emergence of multiple minima of the cost function as $K$ increases. This
is corroborated by the absence of this behavior in the "more unimodal" L95 model, cf.~panel (d), where the three DA-based approaches give indistinguishable estimates.

The numerical results so far pertain to the comparison of the four methods of evaluating the CME integral in Eq.~(\ref{CME}). One can presume that their accuracy depends strongly on the accuracy of the DA-based scheme on which they rely, leaving IS, which suffers from under-sampling and lack of resampling, as the least accurate option. Nevertheless, trying to achieve an overall ranking of DA schemes may be difficult, since different methods may be optimal for specific tasks and under specific circumstances. Such differences in method performance for model evidence are illustrated here by application to the L63 model vs. the L95 one, with the former exhibiting a bimodal attractor.

\section{Applications \label{sec:apply}}%% Illustration}

We briefly discuss now two possible applications of our proposed schemes; both are straightforward to describe and implement based on the results obtained in section~\ref{sec:results}.

\subsection{Parameter estimation}

Let us assume now that the true values of the forcing parameters, {\it i.e.} $\lambda_1$ in L63 and $F_1$ in L95, are unknown, but that we have access to a time series of observations for each one of the two models. Many methods for parameter estimation are available \citep[{e.g.,}][]{Ghil97, Kondras.et.al.08, Kan09, carrassi2011state} but the present maximum likelihood estimation approach is relatively new. A recent review on parameter estimation for the geosciences can be found in \citet{bocquet2015c}.

By maximizing the CME for the parameter of interest, we can use the observations on the model state to evaluate the unknown forcing. In addition, we will apply the standard likelihood ratio approach to derive confidence intervals on the forcing estimates obtained by our CME approach. %\cm{Why doesn't a Monte Carlo approach yield CIs directly? Pls. explain. AH: I don’t really see the connection with a MC approach. In the classic max likelihood estimation approach used here, there is no need for MC to estimate CI. The estimator and its uncertainty are directly  obtained from the likelihood and its usual asymptotic properties (which Bayesians typically don’t trust that much but we don’t want to get involve into this here).}

The results of this approach to parameter estimation are also available in Figs.~\ref{FIG6}(a,b). With $K=10$, the four methods yield an unbiased estimate of the forcing parameter in both the L63 (panel (a)) and the L95 (panel (b)) model, since the CME curves in both panels reach their maxima at the correct parameter value.

The methods differ, however, in the curvature of each of the CME curves at their respective maxima, and thus on the corresponding confidence intervals. More precisely, the less accurate methods yield a higher curvature than the most accurate ones, and thus lower estimates of uncertainty. 
Indeed, the variance of a maximum likelihood estimator is driven by the inverse of the second derivative of the negative log likelihood, asymptotically \citep[{e.g.,}][]{millar2011maximum}.
Accordingly, less accurate methods appear to still be able to yield unbiased parameter estimates, but to underestimate uncertainty. Furthermore, increasing the length $K$ of the evidencing window in Figs.~\ref{FIG6}(c,d) does not affect the position of the maxima and so the estimates remain unbiased, but it appears to increase the gap in CME estimates for parameter values that differ from the truth. Thus increasing $K$ reduces the uncertainty estimates, as expected.

\subsection{Causal attribution of climate-related events}

Providing causal assessments on episodes of extreme weather or unusual climate conditions has become an important topic in the climate sciences over the last decade \citep{Solomon2007climate, Stocker2013ipcc}. Its importance arises from the multiple needs for public dissemination, litigation in a legal context, adaptation to climate change or simply improvement of the science associated with these events \citep{Stott13}. 

The conventional approach to event attribution so far consists in comparing two probabilities, $p_1$ and $p_0$, and in computing the so-called {\it fraction of attributable risk} or $F_{\rm AR}$ of the event under study, where $F_{\rm AR} = 1 - p_0/p_1$. Here $p_1$ is the probability of occurrence of the event in a model $\mathcal{M}_1$ representing the observed climatic conditions, which simulates the real world, referred to as {\em factual}, while $p_0$ is the probability of occurrence of the event in a second model $\mathcal{M}_0$ that represents this time the alternative world that might have occurred had the forcing of interest been absent, referred to as {\em counterfactual}. In the conventional approach, $p_1$ and $p_0$ are calculated by running an ensemble of simulations of each one of the two models, $\mathcal{M}_1$ and $\mathcal{M}_0$, which is computationally quite costly.

Recently, \cite{Hannart-et-al-2016} have shown that this conventional approach can be improved upon by applying DA to derive the CME of a series of observations of the given event for these two alternative models, $\mathcal{M}_1$ and $\mathcal{M}_0$. In both cases, the CME is estimated for the same sequence of meteorological observations of the event. If the value of the factual CME
substantially exceeds that of the counterfactual CME for the event under scrutiny, then it is possible to conclude that
the forcing of interest has had a causal influence on the event.

\begin{figure}
\centering
\includegraphics[height=16pc,width=21pc]{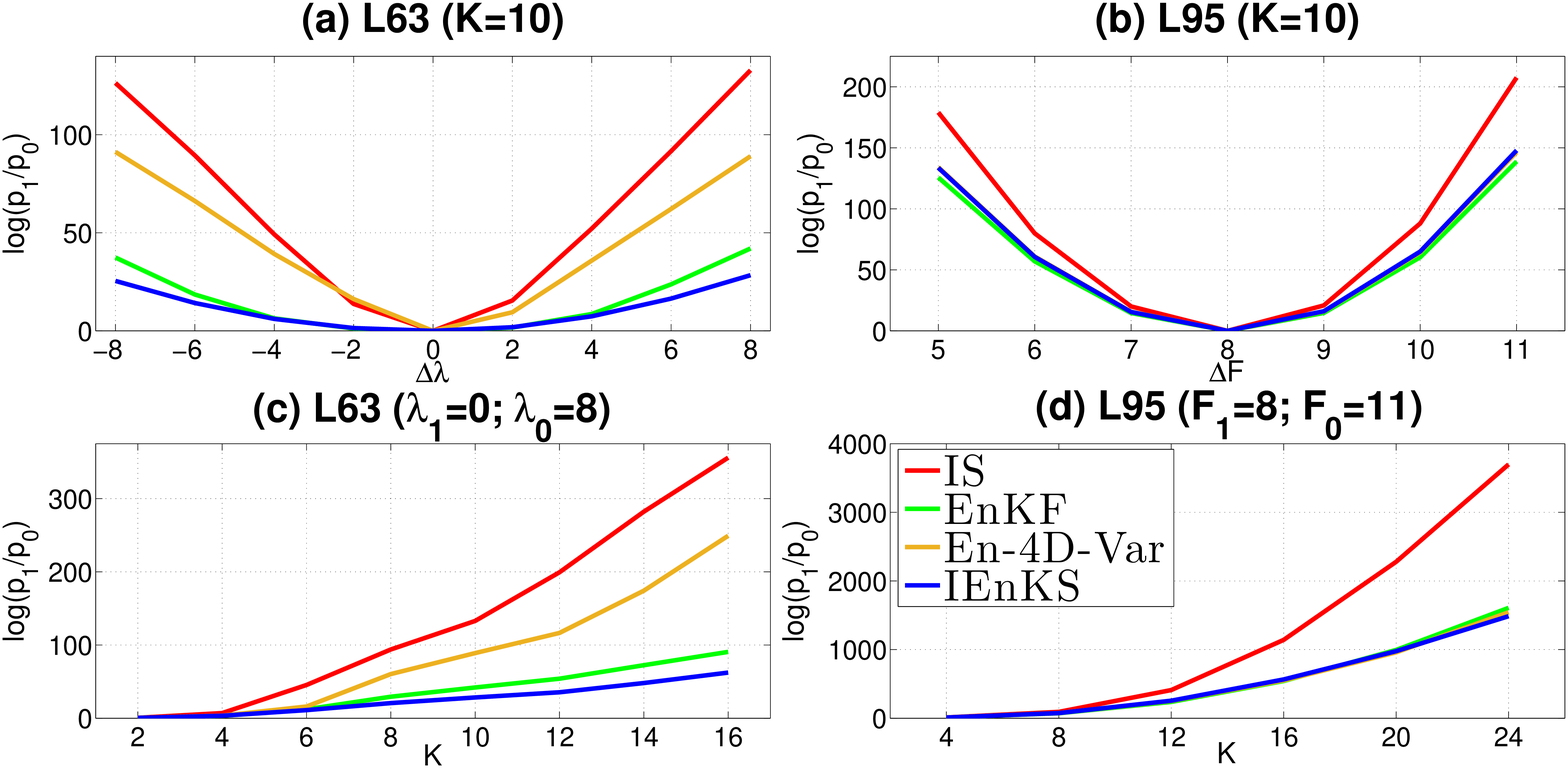}
\caption{Same as Fig.~\ref{FIG6} but for the mean discriminating power.}
\label{FIG7}
\end{figure}

\cite{Hannart-et-al-2016} made use of the EnKF in their DA-based approach to event attribution. Their proposed approach is further investigated here by using the four methods discussed so far to compute model evidence; the effectiveness of these methods is evaluated by calculating the {\it discriminating power} of each of them. This metric is defined as the ability to discriminate whether the observed sequence of observations is more or less likely to occur in the factual rather than in the counterfactual world, and it is obtained as the ratio between the factual and counterfactual model evidence, $p_1/p_0$. Figure~\ref{FIG7} shows the average of the logarithm $\log(p_1/p_0)$ of the discriminating power for L63 and L95, as a function of the counterfactual model's forcing and of the evidencing window length $K$.

%%%%%%%%%%%%%%%%%%%%%%%%%%%%%%%%%%%%%%%%%%%%%%%%%%%%%%%%%%%%%%%%%%%%%%%%
%%%%%%%%%%%%%%%%%%%%%%%%%%%%%%%%%%%%%%%%%%%%%%%%%%%%%%%%%%%%%%%%%%%%%%%%

As expected, the discriminating power increases monotonically as the difference between the factual and the counterfactual forcing increases --- {\it i.e.}, as the cause to whom the event is to be attributed is more salient --- as well as with the length $K$ of the evidencing window. Indeed, the larger $K$ is the more time one can observe the difference between the models.

In agreement with our previous results on the CME, Figs.~\ref{FIG7}(b,d) show that the three DA-based methods provide similar results for the L95 model, while IS systematically overestimates the discriminating power. This means that, if IS is used in an attribution application, it will exaggerate the responsibility of the forcing under scrutiny. The DA-based methods offer a more plausible way to assess the main cause that might have produced the observed sequence.  

The situation is slightly less clear for the L63 model, cf. panels (a) and (c), although IS still provides the largest CME estimate for any given $K$. But the DA-based methods are affected now differently by the increase in the evidencing window, with En-4D-Var displaying the strongest response, 
as already discussed in relation with Fig.~\ref{FIG6}.

\section{Concluding remarks}\label{sec:concl}

\subsection{Summary}

This study focused on the problem of quantifying the resulting performance of a state inference by estimating the model evidence, {\it i.e.} the marginal likelihood $p(\y_{k:})$, which quantifies the accuracy of a model-to-data fit. 
Model evidence is a natural metric for selection and comparison, and it is relevant to many problems faced by both scientists and practitioners: {\it e.g.} calibrating model parameters, comparing the skill of several candidate models in representing the observed signal, or evidencing the existence of a causal relationship between an external forcing and an observed response. This latter attribution problem, and in particular the development of methods for attribution of climate related events in real or near-real time \citep{Stott13, Hannart-et-al-2016}, was one of the original motivations of this study. 

Deriving model evidence for high-dimensional models and big datasets -- both of which are common in the geosciences -- is usually computationally intractable and the issue can only be solved under simplified assumptions \citep[e.g.,][]{HK01}. 
In this paper we have shown how this task can be carried out efficiently using data assimilation (DA) techniques specifically designed to deal with large numerical models and dataset subject to partially Gaussian assumptions. 

We introduced the original {\it contextual} formulation of the model evidence (CME), $p(\y_{k:1}\vert \y_{0:})$ in Eq. (\ref{CME}). As opposed to the standard formulation, the CME is narrowed down to the present systems condition, and the conditional pdf $p(\x_0\vert\y_{0:})$ is taken as the prior in lieu of the invariant distribution $p(\x)$ on the model attractor, as used in the standard case.

This new prior is not only more informative and easier to compute but it is also the outcome, {\it i.e.} the posterior pdf, of a DA procedure designed for state estimation based on the observations. 
When a forecast-assimilation cycle is routinely and continuously carried out, as it is in NWP centers, this prior is immediately at hand.  

We have then proved that DA can also be used to compute the conditional pdf, $p(\y_{k:1}\vert {\bf x}_{0})$, that enters the CME definition, cf. Eq. (\ref{CME}).
Together with the DA-based CME prior, using this pdf also allows one to fully accomplish the model evidence estimation task in a consistent and routine way by using DA. 

Analytic derivations of the proposed DA-based approach were described for both filtering and smoothing methods. In particular, we presented the Kalman filter (KF), the ensemble Kalman filter (EnKF), the ensemble four-dimensional variational smoother (En-4D-Var) and the iterative ensemble Kalman smoother (IEnKS).

The theoretical formulae for the EnKF, En-4D-Var and IEnKS, along with the importance sampling based on the EnKF, have been compared numerically using two low-order chaotic models, the Lorenz 3-variable (L63) model \citep{Lorenz1963} and the 40-variable (L95) model of \citet{lorenz1998}.
Gauss-Hermite quadrature and a massive Monte Carlo algorithm are used as independent highly accurate reference results for the CME values.

Numerical tests were used to compare the computing methods in terms of the CME time series and their distributions, as well as with respect to their sensitivity to the model forcing and to the length of the evidencing window.   
In general the accuracy of the estimates increased with the sophistication of the DA method. Thus the IEnKS yielded the best results and the IS the worst ones. 

In the comparison of method skill for the two models, the IEnKS also performed better for the L63 model, which possesses a bimodal attractor. 
In fact the quasi-static approximation employed by the IEnKS helps preventing the appearance of multiple minima in the cost function that hampers the performance of En-4D-Var. In these conditions also the EnKF appears to behave generally better than En-4D-Var, and its skill is the closest to that of the IEnKS.  
On the other hand, in a weakly nonlinear regime as represented here by the L95 model, the accuracy of the DA-based estimates of the CME appears connected to the level of sophistication of the DA method, with the two smoothers, En-4D-Var and IEnKS, performing better than the filter (EnKF).

We have also considered two applications of estimating the model evidence -- namely to parameter estimation and to causal attribution of climate-related events -- and have studied the potential of the proposed DA-based approach for these purposes. Results have shown that DA-based model evidence (i) can be efficiently used to estimate unknown model parameters, along with the associated uncertainty; and (ii) it is able to discriminate between two competing models and, therewith, to correctly attribute an observed event to one of the two. In both of these applications the IEnKS-based CMEs provide the best performance. Nonetheless the EnKF appears as a good compromise between accuracy and ease of implementation, and its reduced computational cost compared to IEnKS makes it a suitable option for CME computation in a realistic setting.   

\subsection{Future directions}

The next step in applying the present DA-based approach to more realistic models and observational scenarios is to consider climate models of intermediate complexity and incomplete and unevenly distributed observations. 
This application-oriented research activity has to be supported and accompanied by two theoretical lines of investigations, namely the extension of the present results (i) in the presence of model error, and (ii) in conjunction with spatial localization techniques. 

Taking into account model error is crucial for obtaining an accurate and robust estimate of the marginal likelihood, and this requirement becomes even more stringent when model evidence is intended for use in model discrimination or selection. Model error can, in fact, masks the difference between two models under-scrutiny, whether one of the two is correct or not. This effect is an issue for both, standard and DA-based, methods, but particularly for the latter as the accuracy of the DA outcome will naturally depend on the model accuracy, as discussed in \citet[][Fig. 4d]{Hannart-et-al-2016}. 
Methods to incorporate model error in DA procedures for state and parameter estimation have been the subject of an intense, and still active, stream of research \citep[e.g.][]{dee1995line, harlim2013model,raanes2015extending}.
The use of these methods and the study of their adaptation to DA-based CME estimation is definitely worth addressing. 

The applications of the proposed DA-based approach  to large dimensional, spatially extended, dynamical systems necessitate the implementation of localization strategies, commonly used in ensemble-based DA in the geosciences with the aim of compensating for the sampling errors that arise from the use of an insufficiently large ensemble; see \citet{sakov2011relation} and references therein. The extension of the formulae described here to compute CME in conjunction with localization is a central theme among the authors' current lines of investigation.

\section*{Acknowledgments}
We acknowledge discussions with the other members of the DADA team (Philippe Naveau, Aur\'elien Ribes, Juan Ruiz and Manuel Pulido) and with Francis Zwiers and Ted Sheperd on detection and attribution aspects. This research was supported by grant DADA from the Agence Nationale de la Recherche (ANR, France: All authors), and by Grant N00014-16-1-2073 of the Multidisciplinary University Research Initiative (MURI) of the US Office of Naval Research (MG). A. Carrassi has been funded by the Nordic Centre of Excellence EmblA of the Nordic Countries Research Council, NordForsk. 

\section*{Appendix \\ Computation of contextual model evidence using Gauss-Hermite quadrature}

All quadrature methods approximate an integral as a weighted sum of the integrand's values, evaluated at a finite set of well-specified points called nodes.
Gaussian quadrature approximates the integral of an unknown function $f(x)$ over a specified domain $\mathcal{D}$ with a known weighting kernel $\psi(x)$. If the function $f(x)$ is well approximated by a polynomial of order $2m-1$, then a quadrature with $m$ nodes suffices for a good estimate of the integral, according to
\begin{equation}
\tag{A1}
\label{G-Quad}
\int_{\mathcal{D}} \! \mathrm{d}xf(x)\psi(x)\approx\sum_{i=1}^mw_if(x_i)  .
\end{equation}

The nodes $x_i$ and weights $w_i$ are uniquely determined by the choice of the domain $\mathcal{D}$ and the weighting kernel $\psi(x)$, which in turn determines the type of quadrature. The locations of the nodes $\{x_i: i = 1, \ldots, m \}$ are given by the roots of the polynomial of order $m$ in the sequence of orthonormal polynomials $\{\pi_j(x): j = 1, \ldots, m \}$, according to the scalar product $(\pi_j\vert\pi_k)= \int_{\mathcal{D}} \! \mathrm{d}x\, \pi_j(x)\pi_k(x)=\delta_{jk} $, and the weights are computed once the roots are known.

When the integration domain is the entire real axis and the kernel is given by a Gaussian function, the quadrature method is known as the {\it Gauss-Hermite} method, since it involves the orthogonal Hermite polynomials, and it can be written as
\begin{equation}
\tag{A2}
\label{HG-Quad}
\int_{\mathbb R} \! \mathrm{d}xf(x)\exp(-x^2) \simeq \sum_{i=1}^mw_if(x_i) \,,
\end{equation}
with $x_i$  being the roots of the Hermite polynomials ${\mathcal He}_m$ of degree $m$, and the weights $w_i$ being given by
\begin{equation}
\tag{A3}
\label{HG-Quad-W}
w_i = \frac{2^{m-1}m!\sqrt{\pi}}{ m^2\left\{ {\mathcal He}_{m-1}(x_i)\right\}^2} \qquad i=1, \ldots ,m \, .
\end{equation}    
Moreover, when the kernel is given by a normal distribution, $\psi = \mathcal{N}(\overline{x},\sigma^2)$, the Gauss-Hermite approximation becomes
\begin{align*}
\label{HG-Quad-1dG}
\int_{\mathbb R} \! \mathrm{d}x \frac{1}{\sigma\sqrt{2\pi}}\exp\left\{-\frac{\(x-\overline{x}\)^2}{2\sigma^2}\right\}f(x) \simeq \nn 
\frac{1}{\sqrt{\pi}} \sum_{i=1}^mw_if\(\sqrt{2}\sigma x_i + \overline{x}\) \, .  
\tag{A4}
\end{align*}
This formula can be obtained from Eq.~(\ref{HG-Quad}) after a change of variable and integration by substitution. Gauss-Hermite quadrature is of key importance in many areas of applied science, including statistics and finance, and is described in many textbooks on numerical analysis \citep[e.g.,][]{Num_Recipes}.  

Let us recall the general expression of the CME in Eq. (\ref{CME}):
\begin{equation}
\tag{A5}
\label{CME-App}
 p(\y_{K:1}\vert\y_{0:}) = \int  \! \mathrm{d}\x_{0} \, p(\y_{K:1}\vert {\bf x}_{0})p({\bf x}_{0}\vert \y_{0:}).
\end{equation}

In our numerical experiments of section~\ref{sec:results}, the prior density, {\it i.e.} the kernel in Eq.~(\ref{CME-App}), is the multivariate Gaussian posterior obtained from the EnKF applied to the underlying factual trajectory, and it reads
\begin{equation}
\tag{A6}
\label{Ker}
p({\bf x}_{0}\vert \y_{0:}) = \frac{\exp \( -\frac{1}{2} \left\| \x_0 - \barx_0 \right\|^2_{\bP_0^{\mathrm{f}}} \)   }{\sqrt{(2\pi)^M \left| \bP_0^{\mathrm{f}} \right|}} \, .
\end{equation}
Here, the forecast error covariance matrix $\bP_0^{\mathrm{f}}$ is given in terms of the forecast perturbation matrix, $\bP_0^{\mathrm{f}} = \bX_0\bX_0^\T$, as described in section~\ref{ssec:setting}.

We consider the case $N\ge M+1$, for which $\bP_0$ is almost surely non-singular and which applies in the numerical experiments with the L63 model.
The choice of a Gaussian kernel allows for the use of Gauss-Hermite quadrature, and the function $f(x)$ in Eq.~(\ref{HG-Quad-1dG}) is the likelihood $p(\y_{K:1}\vert {\bf x}_{0})$. Given Gaussian observational errors with covariance $\bR$, $f(x)$ should be replaced with
\begin{equation}
\tag{A7}
\label{LikL}
p(\y_{K:1}\vert {\bf x}_{0}) = \frac{\exp \( -\frac{1}{2} \sum_{k=1}^K \left\| \y_k - \mathcal{H}_k\circ{\mathcal M}_{k:0}(\x_0) \right\|^2_{\bR_k}\)}{\sqrt{(2\pi)^{Kd}\prod_{k=1}^K \left|\bR_k\right|}}.
\end{equation}

We now consider the multivariate extension of the Gauss-Hermite quadrature Eq.~(\ref{HG-Quad-1dG}). In order to make the sampling more efficient, it is convenient to use the principal axes of variability of the covariance matrix $\bP_0^{\mathrm{f}}$. To this end, let us decompose the perturbation matrix,
\begin{equation}
\tag{A8}
\label{X-SV}
\bX_0 = \bU\bS\bV^\T \, .
\end{equation}
Here, $\bU$ and $\bV$ are the matrices whose columns are the $M$ left and right singular vectors, and $\bS$ is the diagonal matrix of the $M$ singular values.   
We define the operator
\begin{equation}
\tag{A9}
\label{Op}
\bZ = \sqrt{2}\bU^\T\bS\bU  \, .
\end{equation}
By proceeding via integration by substitution, as for the univariate case, the multivariate Gauss-Hermite quadrature method can be shown to be given by
\begin{align}
\label{HG-Quad-MdG}
&\int_{{\mathbb R}^M}  \! \mathrm{d}\x \frac{1}{\sqrt{(2\pi)^M\left|\bP_0^{\mathrm{f}}\right| }} \exp\(-\frac{1}{2} \left\| \x - \barx \right\|^2_{\bP_0^{\mathrm{f}}} \) f(\x) \nn 
& \simeq  \sum^m_{i_1=1,i_2=1, \ldots ,i_M=1}\prod_{j=i_1}^{i_M} w_j f(\barx + \bZ\bchi)
\tag{A10}
\end{align}
with $\bchi=(x_{i_1},x_{i_2},...,x_{i_M})$ being the roots of the Hermite polynomials. The final formula to approximate Eq.~(\ref{CME-App}) is obtained by using Eq.~(\ref{LikL}) as $f(\x)$.\\

\bibliographystyle{wileyqj}
\bibliography{library_v6}

\begin{thebibliography}{59}
\providecommand{\natexlab}[1]{#1}
\providecommand{\url}[1]{\texttt{#1}}
\providecommand{\urlprefix}{URL }
\expandafter\ifx\csname urlstyle\endcsname\relax
  \providecommand{\doi}[1]{doi:\discretionary{}{}{}#1}\else
  \providecommand{\doi}{doi:\discretionary{}{}{}\begingroup
  \urlstyle{rm}\Url}\fi

\bibitem[{Balmaseda \emph{et~al.}(2009)Balmaseda, Alves, Arribas, Awaji,
  Behringer, Ferry, Fujii, Lee, Rienecker, Rosati and Stammer}]{Bal09}
Balmaseda MA, Alves OJ, Arribas A, Awaji T, Behringer DW, Ferry N, Fujii Y, Lee
  T, Rienecker M, Rosati T, Stammer D. 2009. Ocean initialization for seasonal
  forecasts. \emph{Oceanography Special Issue} \textbf{22}: 154.

\bibitem[{Baum \emph{et~al.}(1970)Baum, Petrie, Soules and Weiss}]{Baum70}
Baum L, Petrie T, Soules G, Weiss N. 1970. A maximization technique occurring
  in the statistical analysis of probabilistic functions of {M}arkov chains.
  \emph{Ann. Math. Stat.} \textbf{41}: 164--171.

\bibitem[{Bengtsson \emph{et~al.}(1981)Bengtsson, Ghil and
  K\"{a}ll\'{e}n}]{Bengtsson81}
Bengtsson L, Ghil M, K\"{a}ll\'{e}n E (eds). 1981. \emph{Dynamic meteorology:
  Data assimilation methods}. Springer-Verlag, New York/Heidelberg/Berlin.

\bibitem[{Bhend \emph{et~al.}(2012)Bhend, Franke, Folini, Wild and
  Br\'{o}nnimann}]{Bhe12}
Bhend J, Franke J, Folini D, Wild M, Br\'{o}nnimann S. 2012. An ensemble-based
  approach to climate reconstructions. \emph{Clim. Past.} \textbf{8}: 963--976.

\bibitem[{Blayo \emph{et~al.}(2015)Blayo, Bocquet, Cosme and
  Cugliandolo}]{blayo2015}
Blayo E, Bocquet M, Cosme E, Cugliandolo LF (eds). 2015. \emph{Advanced data
  assimilation for geosciences}, Oxford University Press. Lecture Notes of the
  Les Houches School of Physics: Special Issue, June 2012.

\bibitem[{Bocquet(2015)}]{bocquet2015c}
Bocquet M. 2015. An introduction to inverse modelling and parameter estimation
  for atmosphere and ocean sciences. In: \emph{Advanced data assimilation for
  geosciences}, Blayo {\'E}, Bocquet M, Cosme E, Cugliandolo LF\ (eds). Oxford
  University Press: Les {H}ouches school of physics, pp. 461--493.

\bibitem[{Bocquet(2016)}]{bocquet2016}
Bocquet M. 2016. Localization and the iterative ensemble {K}alman smoother.
  \emph{Q J Roy. Meteor. Soc.} \textbf{142}: 1075--1089.

\bibitem[{Bocquet \emph{et~al.}(2010)Bocquet, Pires and Wu}]{Bocquet2010}
Bocquet M, Pires C, Wu L. 2010. {Beyond Gaussian Statistical Modeling in
  Geophysical Data Assimilation}. \emph{Mon. Weather Rev.} \textbf{138}:
  2997--3023.

\bibitem[{Bocquet and Sakov(2013)}]{bocquet2013}
Bocquet M, Sakov P. 2013. Joint state and parameter estimation with an
  iterative ensemble {K}alman smoother. \emph{Nonlinear Proc. Geoph.}
  \textbf{20}: 803--818.

\bibitem[{Bocquet and Sakov(2014)}]{bocquet2014}
Bocquet M, Sakov P. 2014. {An iterative ensemble Kalman smoother}. \emph{Q J
  Roy. Meteor. Soc.} \textbf{140}: 1521--1535.

\bibitem[{Carrassi and Vannitsem(2010)}]{CV2010}
Carrassi A, Vannitsem S. 2010. {Accounting for model error in variational data
  assimilation: A deterministic formulation}. \emph{Mon. Weather Rev.}
  \textbf{138}: 3369--3386.

\bibitem[{Carrassi and Vannitsem(2011)}]{carrassi2011state}
Carrassi A, Vannitsem S. 2011. State and parameter estimation with the extended
  kalman filter: an alternative formulation of the model error dynamics.
  \emph{Q J Roy. Meteor. Soc.} \textbf{137}(655): 435--451.

\bibitem[{Carrassi \emph{et~al.}(2008)Carrassi, Vannitsem and
  Nicolis}]{CVN2008}
Carrassi A, Vannitsem S, Nicolis C. 2008. Model error and sequential data
  assimilation: A deterministic formulation. \emph{Q J Roy. Meteor. Soc.}
  \textbf{134}: 1297--1313.

\bibitem[{Carson \emph{et~al.}(2015)Carson, Crucifix, Preston and
  Wilkinson}]{Carson_et_al_2015}
Carson J, Crucifix M, Preston S, Wilkinson RD. 2015. Bayesian model selection
  for the glacial-interglacial cycle. \emph{arXiv preprint arXiv:1511.03467} .

\bibitem[{Chekroun \emph{et~al.}(2011)Chekroun, Simonnet and
  Ghil}]{Chekroun_et_al_2011}
Chekroun M, Simonnet E, Ghil M. 2011. {Stochastic climate dynamics: random
  attractors and time-dependent invariant measures}. \emph{Physica D}
  \textbf{240}: 1685--1700.

\bibitem[{Crisan and Doucet(2002)}]{Cris02}
Crisan D, Doucet A. 2002. A survey of convergence results on particle filtering
  for practitioners. \emph{IEEE T. Signal Proces.} \textbf{50}: 736--746.

\bibitem[{Dee(1995)}]{dee1995line}
Dee DP. 1995. On-line estimation of error covariance parameters for atmospheric
  data assimilation. \emph{Mon. Weather Rev.} \textbf{123}(4): 1128--1145.

\bibitem[{Del~Moral(2004)}]{DelMoral_2004}
Del~Moral P. 2004. \emph{{Feynman-Kac formulae: genealogical and interacting
  particle systems with applications}}. Springer: New York.

\bibitem[{Dijkstra(2013)}]{Dijkstra_2013}
Dijkstra H. 2013. \emph{{ Nonlinear Climate Dynamics}}. Cambridge University
  Press: Cambridge.

\bibitem[{Elsheikh \emph{et~al.}(2014{\natexlab{a}})Elsheikh, Hoteit and
  Wheeler}]{Els14a}
Elsheikh A, Hoteit I, Wheeler M. 2014{\natexlab{a}}. Efficient bayesian
  inference of subsurface flow models using nested sampling and sparse
  polynomial chaos surrogates. \emph{Comput. Methods Appl. Mech. Engrg.}
  \textbf{269}: 515--537.

\bibitem[{Elsheikh \emph{et~al.}(2014{\natexlab{b}})Elsheikh, Wheeler and
  Hoteit}]{Els14b}
Elsheikh A, Wheeler M, Hoteit I. 2014{\natexlab{b}}. Hybrid nested sampling
  algorithm for bayesian model selection applied to inverse subsurface flow
  problems. \emph{J. Comput. Phys.} \textbf{258}: 319--337.

\bibitem[{Evans and Swartz(1995)}]{Evans_Swartz_1995}
Evans M, Swartz T. 1995. {Methods for approximating integrals in statistics
  with special emphasis on Bayesian integration problems}. \emph{Stat. Sci.}
  \textbf{10}: 254--272.

\bibitem[{Evensen(2009)}]{evensen2009}
Evensen G. 2009. \emph{{D}ata {A}ssimilation: {T}he {E}nsemble {K}alman
  {F}ilter}. Springer-Verlag/Berlin/Heildelberg, second edn.

\bibitem[{Ghil(1997)}]{Ghil97}
Ghil M. 1997. Advances in sequential estimation for atmospheric and oceanic
  flows. \emph{J. Meteorol. Soc. Jpn.} \textbf{75}: 289--304.

\bibitem[{Ghil \emph{et~al.}(2008)Ghil, Chekroun and Simonnet}]{GCS08}
Ghil M, Chekroun MD, Simonnet E. 2008. Climate dynamics and fluid mechanics:
  Natural variability and related uncertainties. \emph{Physica D} \textbf{237}:
  2111--2126.

\bibitem[{Ghil and Malanotte-Rizzoli(1991)}]{GMR91}
Ghil M, Malanotte-Rizzoli P. 1991. Data assimilation in meteorology and
  oceanography. \emph{Adv. Geophys.} \textbf{33}: 141--266.

\bibitem[{Grewal and Andrews(2001)}]{Grewal01}
Grewal M, Andrews A. 2001. \emph{Kalman filtering : Theory and practice using
  matlab}. Wiley-Interscience, 2 edn.

\bibitem[{Hannart \emph{et~al.}(2016)Hannart, Carrassi, Bocquet, Ghil, Naveau,
  Pulido, Ruiz and Tandeo}]{Hannart-et-al-2016}
Hannart A, Carrassi A, Bocquet M, Ghil M, Naveau P, Pulido M, Ruiz J, Tandeo P.
  2016. Dada: data assimilation for the detection and attribution of weather
  and climate-related events. \emph{Climatic Change} \textbf{136}(2): 155--174.

\bibitem[{Harlim(2013)}]{harlim2013model}
Harlim J. 2013. Model error in data assimilation. \emph{arXiv preprint
  arXiv:1311.3579} .

\bibitem[{Hunt \emph{et~al.}(2007)Hunt, Kostelich and Szunyogh}]{hunt2007}
Hunt B, Kostelich EJ, Szunyogh I. 2007. Efficient data assimilation for
  spatiotemporal chaos: {A} local ensemble transform {K}alman filter.
  \emph{Physica D} \textbf{230}: 112--126.

\bibitem[{H\"urzeleri and K\"unsch(2001)}]{HK01}
H\"urzeleri M, K\"unsch H. 2001. Approximation and maximising the likelihood
  for a general state-space model. In: \emph{Sequential Monte Carlo Methods in
  Practice}, Springer-Verlag: New York.

\bibitem[{Jazwinski(1970)}]{jazwinski1970}
Jazwinski AH. 1970. \emph{Stochastic processes and filtering theory}. Academic
  Press, New-York.

\bibitem[{Kalman(1960)}]{kalman1960}
Kalman RE. 1960. A new approach to linear filtering and prediction problems.
  \emph{Journal of Fluids Engineering} \textbf{82}: 35--45.

\bibitem[{Kalnay(2002)}]{Kalnay2002}
Kalnay E. 2002. \emph{{Atmospheric Modeling, Data Assimilation and
  Predictability}}. Cambridge University Press: Cambridge.

\bibitem[{Kantas \emph{et~al.}(2009)Kantas, Doucet, Singh and
  Maciejowski}]{Kan09}
Kantas N, Doucet A, Singh S, Maciejowski J. 2009. An overview of sequential
  {Monte Carlo} methods for parameter estimation. In: \emph{General State-Space
  Models}, IFAC System Identification.

\bibitem[{Kondrashov \emph{et~al.}(2008)Kondrashov, Sun and
  Ghil}]{Kondras.et.al.08}
Kondrashov D, Sun Cj, Ghil. 2008. Data assimilation for a coupled
  ocean-atmosphere model. {Part II: Parameter estimation}. \emph{Mon. Weather
  Rev.} \textbf{136}: 5062--5076.

\bibitem[{Liu and Pierce(1994)}]{Liu_Pierce_1994}
Liu Q, Pierce D. 1994. {A Note on Gauss-Hermite Quadrature}. \emph{Biometrika}
  \textbf{81}: 624--629.

\bibitem[{Lorenc(2013)}]{Lorenc2013}
Lorenc A. 2013. \emph{{ Recommended Nomenclature for EnVar Data Assimilation
  Methods }}. Available
  at:http://www.wcrp-climate.org/WGNE/BlueBook/2013/individual-articles/01\_Lorenc\_Andrew\_EnVar\_nomenclature.pdf.

\bibitem[{Lorenz(1963)}]{Lorenz1963}
Lorenz E. 1963. {Deterministic non-periodic flow}. \emph{J. Atmos. Sci.}
  \textbf{20}: 130--141.

\bibitem[{Lorenz and Emanuel(1998)}]{lorenz1998}
Lorenz EN, Emanuel KA. 1998. Optimal sites for supplementary weather
  observations: simulation with a small model. \emph{J. Atmos. Sci.}
  \textbf{55}: 399--414.

\bibitem[{Millar(2011)}]{millar2011maximum}
Millar RB. 2011. \emph{Maximum likelihood estimation and inference: with
  examples in r, sas and admb}, vol. 111. John Wiley \& Sons.

\bibitem[{Miller \emph{et~al.}(1994)Miller, Ghil and Gauthiez}]{Milleretal94}
Miller R, Ghil M, Gauthiez F. 1994. Advanced data assimilation in strongly
  nonlinear dynamical systems. \emph{J. Atmos. Sci.} \textbf{{\bf 51}}:
  1037--1056.

\bibitem[{Palmer(1999)}]{Palmer1999}
Palmer T. 1999. {A non-linear dynamical perspective on climate prediction}.
  \emph{J. Climate} \textbf{12}: 575--591.

\bibitem[{Pearl(2000)}]{Pearl00}
Pearl J. 2000. \emph{Causality: Models, reasoning and inference}. Cambridge
  Univ. Press, Cambridge, UK, and New York, NY, USA.

\bibitem[{Pires \emph{et~al.}(1996)Pires, Vautard and Talagrand}]{pires1996}
Pires C, Vautard R, Talagrand O. 1996. On extending the limits of variational
  assimilation in nonlinear chaotic systems. \emph{Tellus A} \textbf{48}:
  96--121.

\bibitem[{Pitt(2002)}]{Pitt02}
Pitt M. 2002. Smooth particle filters for likelihood evaluation and
  maximisation. \emph{Warwick Economic Research Papers} \textbf{651}.

\bibitem[{Press \emph{et~al.}(1992)Press, Teukolsky, Vetterling and
  Flannery}]{Num_Recipes}
Press WH, Teukolsky SA, Vetterling WT, Flannery BP. 1992. \emph{{ Numerical
  Recipes in Fortran 77}}. Cambridge University Press: Cambridge.

\bibitem[{Raanes \emph{et~al.}(2015)Raanes, Carrassi and
  Bertino}]{raanes2015extending}
Raanes PN, Carrassi A, Bertino L. 2015. Extending the square root method to
  account for additive forecast noise in ensemble methods. \emph{Mon. Weather
  Rev.} \textbf{143}: 3857--3873.

\bibitem[{Reich and Cotter(2015)}]{Reich-Cotter-2015}
Reich S, Cotter C. 2015. \emph{{ Probabilistic Forecasting and Bayesian Data
  Assimilation}}. Cambridge University Press: Cambridge.

\bibitem[{Sakov and Bertino(2011)}]{sakov2011relation}
Sakov P, Bertino L. 2011. Relation between two common localisation methods for
  the enkf. \emph{Computat. Geosci.} \textbf{15}(2): 225--237.

\bibitem[{Solomon \emph{et~al.}(2007)Solomon, Qin, Manning, Chen, Marquis,
  Averyt, Tignor and Miller}]{Solomon2007climate}
Solomon S, Qin D, Manning M, Chen Z, Marquis M, Averyt KB, Tignor M, Miller HL
  (eds). 2007. \emph{{Climate Change 2007. The Physical Science Basis: Working
  Group I Contribution to the Fourth Assessment Report of the IPCC}}. Cambridge
  University Press.

\bibitem[{Stocker \emph{et~al.}(2013)Stocker, Qin, Plattner, Tignor, Allen,
  Boschung, Nauels, Xia, Bex and Midgley}]{Stocker2013ipcc}
Stocker TF, Qin D, Plattner GK, Tignor M, Allen SK, Boschung J, Nauels A, Xia
  Y, Bex B, Midgley BM (eds). 2013. \emph{{IPCC, 2013. Climate Change 2013. The
  Physical Science Basis: Contribution of Working Group I to the Fifth
  Assessment Report of the Intergovernmental Panel on Climate Change}}.
  Cambridge University Press.

\bibitem[{Stott \emph{et~al.}(2013)Stott, Allen, Christidis, Dole, Hoerling,
  Huntingford, Pall, Perlwitz and Stone}]{Stott13}
Stott P, Allen M, Christidis N, Dole R, Hoerling M, Huntingford C, Pall P,
  Perlwitz J, Stone D. 2013. Attribution of weather and climate-related events.
  In: \emph{Climate Science for Serving Society: Research, Modelling and
  Prediction Priorities. G.R. Asrar and J.~W. Hurrell (Eds.)}, Springer.

\bibitem[{Talagrand and Courtier(1987)}]{TalaCour87}
Talagrand O, Courtier P. 1987. Variational assimilation of meteorological
  observations with the adjoint vorticity equation, {I. Theory}. \emph{Q J Roy.
  Meteor. Soc.} \textbf{113}: 1311--1328.

\bibitem[{Tandeo \emph{et~al.}(2015)Tandeo, Pulido and Lott}]{Tand15}
Tandeo P, Pulido M, Lott F. 2015. Offline parameter estimation using {EnKF} and
  maximum likelihood error covariance estimates: Application to a subgrid-scale
  orography parametrization. \emph{Q J Roy. Meteor. Soc.} \textbf{141}:
  383--395.

\bibitem[{Wiener(1949)}]{Wiener49}
Wiener N. 1949. \emph{Extrapolation, interpolation and smoothing of stationary
  time series, with engineering applications}. M.I.T. Press: Cambridge, MA.

\bibitem[{Winiarek \emph{et~al.}(2012)Winiarek, Bocquet, Saunier and
  Mathieu}]{winiarek2012}
Winiarek V, Bocquet M, Saunier O, Mathieu A. 2012. Estimation of errors in the
  inverse modeling of accidental release of atmospheric pollutant:
  {A}pplication to the reconstruction of the cesium-137 and iodine-131 source
  terms from the {F}ukushima {D}aiichi power plant. \emph{J. Geophys. Res.}
  \textbf{117}: D05\,122.

\bibitem[{Winiarek \emph{et~al.}(2011)Winiarek, Vira, Bocquet, Sofiev and
  Saunier}]{Winiarek-et-al-2011}
Winiarek V, Vira J, Bocquet M, Sofiev M, Saunier O. 2011. {Towards the
  operational estimation of a radiological plume using data assimilation after
  a radiological accidental atmospheric release}. \emph{Atmos. Environ.}
  \textbf{45}: 2944--2955.

\bibitem[{Zhang \emph{et~al.}(2012)Zhang, Bocquet, Mallet, Seigneur and
  Baklanov}]{zhang2012}
Zhang Y, Bocquet M, Mallet V, Seigneur C, Baklanov A. 2012. Real-time air
  quality forecasting, part {II}: State of the science, current research needs,
  and future prospects. \emph{Atmos. Environ.} \textbf{60}: 656--676.

\end{thebibliography}

\end{document}